\documentclass[useAMS,usenatbib, letterpaper]{mn2e}

\usepackage{natbib}
\usepackage{graphicx,graphics}
\usepackage{amsmath} 
\usepackage{amssymb}
\def\up{_{\rm u}}
\def\down{_{\rm d}}

\title[]{Using double radio relics to constrain galaxy cluster mergers:  A model of double radio relics in CIZA~J2242.8+5301}
\author[R.~J. van Weeren, M. Br\"uggen, H.~J.~A. R\"ottgering and M. Hoeft]{R.~J. van Weeren$^{1}$\thanks{E-mail:
rvweeren@strw.leidenuniv.nl}, M. Br\"uggen$^{2}$, H.~J.~A. R\"ottgering$^{1}$ and M. Hoeft$^{3}$ \\ \\
$^{1}$Leiden Observatory, Leiden University, P.O. Box 9513, NL-2300 RA Leiden, The Netherlands\\
$^{2}$Jacobs University Bremen, P.O. Box 750561, 28725 Bremen, Germany \\
$^{3}$Th\"uringer Landessternwarte Tautenburg, Sternwarte 5, 07778, Tautenburg, Germany}
\begin{document}

\date{ } 

\pagerange{\pageref{firstpage}--\pageref{lastpage}} \pubyear{2011}

\maketitle

\def\aj{{AJ}}                   
\def\araa{{ARA\&A}}             
\def\apj{{ApJ}}                 
\def\apjl{{ApJ}}                
\def\apjs{{ApJS}}               
\def\ao{{Appl.~Opt.}}           
\def\apss{{Ap\&SS}}             
\def\aap{{A\&A}}                
\def\aapr{{A\&A~Rev.}}          
\def\aaps{{A\&AS}}              
\def\azh{{AZh}}                 
\def\baas{{BAAS}}               
\def\jrasc{{JRASC}}             
\def\memras{\ref@jnl{MmRAS}}            
\def\mnras{{MNRAS}}             
\def\pra{\ref@jnl{Phys.~Rev.~A}}        
\def\prb{\ref@jnl{Phys.~Rev.~B}}        
\def\prc{\ref@jnl{Phys.~Rev.~C}}        
\def\prd{\ref@jnl{Phys.~Rev.~D}}        
\def\pre{\ref@jnl{Phys.~Rev.~E}}        
\def\prl{\ref@jnl{Phys.~Rev.~Lett.}}    
\def\pasp{{PASP}}               
\def\pasj{{PASJ}}               
\def\qjras{\ref@jnl{QJRAS}}             
\def\skytel{\ref@jnl{S\&T}}             
\def\solphys{\ref@jnl{Sol.~Phys.}}      
\def\sovast{\ref@jnl{Soviet~Ast.}}      
\def\ssr{\ref@jnl{Space~Sci.~Rev.}}     
\def\zap{\ref@jnl{ZAp}}                 
\def\nat{{Nature}}              
\def\iaucirc{\ref@jnl{IAU~Circ.}}
\def\aplett{\ref@jnl{Astrophys.~Lett.}}
\def\apspr{\ref@jnl{Astrophys.~Space~Phys.~Res.}}
\def\bain{\ref@jnl{Bull.~Astron.~Inst.~Netherlands}}
\def\fcp{\ref@jnl{Fund.~Cosmic~Phys.}}
\def\gca{\ref@jnl{Geochim.~Cosmochim.~Acta}}
\def\grl{\ref@jnl{Geophys.~Res.~Lett.}}
\def\jcp{\ref@jnl{J.~Chem.~Phys.}}      
\def\jgr{\ref@jnl{J.~Geophys.~Res.}}    
\def\jqsrt{\ref@jnl{J.~Quant.~Spec.~Radiat.~Transf.}}
\def\memsai{\ref@jnl{Mem.~Soc.~Astron.~Italiana}}
\def\nphysa{\ref@jnl{Nucl.~Phys.~A}}
\def\physrep{{Phys.~Rep.}}
\def\physscr{\ref@jnl{Phys.~Scr}}
\def\planss{\ref@jnl{Planet.~Space~Sci.}}
\def\procspie{\ref@jnl{Proc.~SPIE}}
\let\astap=\aap
\let\apjlett=\apjl
\let\apjsupp=\apjs
\let\applopt=\ao

\label{firstpage}

\begin{abstract}
Galaxy clusters grow by mergers with other clusters and galaxy groups. These mergers create shock waves within the intracluster medium (ICM) that can accelerate particles to extreme energies. In the presence of magnetic fields, relativistic electrons form large regions emitting synchrotron radiation, so-called radio relics. Behind the shock front, synchrotron and inverse Compton (IC) losses cause the radio spectral index to steepen away from the shock front.  In a binary cluster merger, two shock waves are generated which move diametrically outwards along the merger axis. Two radio relics can then form on both sides of the cluster center. An example of such a cluster is CIZA~J2242.8+5301, where very clear spectral steepening in the downstream region is observed. The main relic has a total extent of 1700~kpc, while its width is only 55~kpc. Together with the high observed polarization fraction, this implies the relic is seen very close to edge-on which makes it easier to constrain the merger geometry. Here we present hydrodynamical simulations of idealized binary cluster mergers with the aim of  constraining the merger scenario for this cluster. From our simulations, we find that CIZA~J2242.8+5301 is probably undergoing a merger in the plane of the sky (less then $10^{\circ}$ from edge-on) with a mass ratio ($M_1:M_2$) of about $2:1$, and an impact parameter $\lesssim 400$~kpc. We find that the core passage of the clusters happened about 1~Gyr ago. We conclude that double relics relics can set constraints on the mass ratios, impact parameters, timescales, and viewing geometry of binary cluster mergers, which is particularly useful when detailed X-ray observations are not available. In addition, the presence of large radio relics can be used to constrain the degree of clumping in the outskirts of the ICM, which is important to constrain the baryon fraction, density and entropy profiles, around the virial radius and beyond. We find that the amplitude of density fluctuations, with sizes of $\lesssim 200$~kpc, in the relic in CIZA~J2242.8+5301 is not larger than 30\%.

\end{abstract}


\section{Introduction}

Radio relics are diffuse, steep-spectrum radio sources found in merging galaxy clusters.  They have been divided into three groups \citep{2004rcfg.proc..335K}: radio gischt, radio phoenix and AGN relics. Here we will only be concerned with so-called \emph{radio gischt} relics, these are elongated arc-like radio sources with sizes of up to 2~Mpc \citep[e.g.,][]{1998A&A...332..395E}. They are mostly found in the outskirts of galaxy clusters. Recent observations have given  support to the hypothesis that they trace shock fronts in which particles are accelerated via the diffusive shock acceleration mechanism \citep[DSA;][]{1977DoSSR.234R1306K, 1977ICRC...11..132A, 1978MNRAS.182..147B, 1978MNRAS.182..443B, 1978ApJ...221L..29B, 1983RPPh...46..973D, 1987PhR...154....1B, 1991SSRv...58..259J, 2001RPPh...64..429M}.

A particularly interesting ``class'' of  radio gischt are the so-called double-relics. In this case two relics are diametrically located on both sides of the cluster center 
\citep[e.g., ][]{2009A&A...494..429B, 2009A&A...506.1083V, 2007A&A...463..937V, 2006Sci...314..791B, 1997MNRAS.290..577R, 2010Sci...330..347V, 2011ApJ...727L..25B, 2011A&A...528A..38V, 2011ApJ...736L...8B}. These double relics are very rare, with only about ten known so far. However, most of them were found in the last few years suggesting that more of them await discovery, especially with future deep radio surveys. Large cosmological simulations that include radio emission from shocks, also indicate that double radio relics are more common \citep[e.g.,][]{2011ApJ...735...96S, 2009MNRAS.393.1073B,2008MNRAS.391.1511H}.

Double radio relics are thought to trace outward moving shock waves from a binary cluster merger, which develop after core passage of the two subclusters. In the DSA scenario, synchrotron and IC losses cool the particles at the back of the shock front. If the relics are seen roughly edge-on this creates a radio spectral index gradient towards the cluster center. The line connecting the two relics  should represent the (projected) merger axis of the system.
Thus, double relics in principle can be used to set constraints on the merger timescale, mass ratio, impact parameter and viewing angle. 

An example of a cluster hosting a double radio relic is  CIZA~J2242.8+5301 \citep{2010Sci...330..347V}. 
The cluster is located at $z=0.1921$ and has an X-ray luminosity of $6.8 \times 10^{44}$~ergs$^{-1}$, between 0.1 and 2.4~keV  \citep{2007ApJ...662..224K}.  With this X-ray luminosity the total mass of the cluster is roughly $5.5 \times 10^{14}\rm{M_{\odot}}$ \citep{2009A&A...498..361P}. In \cite{2010Sci...330..347V} we presented Westerbork Synthesis Radio Telescope (WSRT) and Metrewave Radio Telescope (GMRT) observations of this cluster that showed the presence of an impressive double relic system. In the north of the cluster, a large radio relic is found with a total extent of 1700~kpc (we did not include the patch of emission at RA~22$^\mathrm{h}$~42$^\mathrm{m}$~20$^\mathrm{s}$, 
Dec~$+$53\degr~08\arcmin~45\arcsec).
To the south of the cluster center a second fainter relic is found with an extent of 1450~kpc. The relics are located along the major axis of the elongated X-ray emission from the ICM, while the relics themselves are orientated perpendicular to this axis. This configuration is expected for a ``clean'' (i.e., a merger event without much substructure) binary merger event in which shock waves are propagating diametrically outwards to the north and south. Within the shock particles are proposed to be accelerated by DSA. The relics are separated by a distance of 2.8~Mpc. The northern relic is strongly polarized at 4.9~GHz, with a polarisation fraction of $\sim 50\%$. The polarization magnetic field vectors (corrected for the effect of Faraday rotation) are parallel to the major axis of the relic. The magnetic fields are thus mainly orientated within the plane of the shock wave and the relic must be seen close to edge-on \citep{1998A&A...332..395E}. We also observe clear steepening of the radio spectrum (with the spectral index changing from $-0.6 \sim -0.8$ to more than $-2.0$ between 0.61 and 2.3~GHz) in the direction of the cluster center, across the width of the relic. Between the two relics additional diffuse emission is found, which is also seen in a deep 150~MHz GMRT image \citep{2011arXiv1101.5161V}.

However, there are still quite a few puzzles that surround radio relics, such as:

\begin{itemize}
\item Which processes accelerate electrons so efficiently at relatively low Mach number ($M\sim 2-4$) shocks?
\item What produces the magnetic fields inside relics?
\item Why do some relics have very sharp edges while others appear very fuzzy?
\item Under which conditions do relics form? When do we see single and when double relics?
\end{itemize}

Many of these questions touch physical processes that are poorly understood, such as diffusive shock acceleration and magnetogenesis in collisionless shocks, and relate to regions of the cosmos of which we know very little. At the same time, outskirts of galaxy clusters have attracted considerable interest for cosmology as they are governed primarily by simple gas physics and gravity and much less by complicated processes such as radiative cooling, star formation and AGN feedback. However, information about the periphery of galaxy clusters is scarce. Recent X-ray observations with Suzaku have measured surface brightness profiles out to the virial radius for a number of nearby clusters \citep{2009PASJ...61.1117B, 2009MNRAS.395..657G, 2009A&A...501..899R, 2010PASJ...62..371H, 2011Sci...331.1576S}. The results indicate significant deviations from extrapolated profiles of the gas mass fraction which are explained by an increased clumpiness of the ICM that occurs on scales of several 100~kpc or smaller. As the cluster outskirts are difficult to access via X-ray observations owing to the low surface brightness at large cluster-centric radii, important information may be derived from radio relics which illuminate the outskirts of merging galaxy clusters. 

In this paper we will describe simulations with the aim of reconstructing the merger event of the cluster CIZA~J2242.8+5301.  In particular we will focus on the mass ratio and impact parameter of the merger and how they can be constrained from the observations and simulations. We further attempt to infer properties of the ICM at distances beyond the virial radius from the observed features of the radio relics, such as the morphology of the emission.

In Sect.~\ref{sec:overview} we will shortly discuss previous work on idealized models of cluster mergers, and in Sect.~\ref{sec:method} we describe  the method and adopted initial conditions. This is followed the results in Sects.~\ref{sec:results} and \ref{sec:substructure}. We end with a discussion and summary in Sects.~\ref{sec:discussion} and \ref{sec:summary}.

\section{Overview: Simulations of galaxy cluster mergers}
\label{sec:overview}
Numerical simulations of idealized binary galaxy cluster mergers were performed by \cite{1993ApJ...407L..53R,1993A&A...272..137S, 1994ApJ...427L..87B, 1994MNRAS.268..953P, 1997ApJS..109..307R, 1998ApJ...496..670R,2000ApJ...538...92R, 2001ApJ...561..621R}.

\cite{1993ApJ...407L..53R} simulated a head-on merger with a mass ratio of 8. The clusters were modeled as King-spheres. They used a hybrid Hydro/N-body code, simulating the hydrodynamical component with the ZEUS-3D code on a nonuniform grid, and the N-body component with the Hernquist treecode \citep{1987ApJS...64..715H}. Their simulation showed the development of a single shock that reached a Mach number ($\mathcal{M}$) of more than 4. In addition the core of the main cluster was heated to $3 \times 10^{8}$~K. The simulated X-ray surface brightness was double peaked and elongated parallel to the merger axis. \cite{1994ApJ...427L..87B} simulated the collision between the {Coma cluster} and the {NGC~4839} galaxy group using the same code as described by \cite{1993ApJ...407L..53R}. 

\cite{1993A&A...272..137S} simulated collisions of clusters using a uniform grid-PPM (piecewise-parabolic method) code for the gas dynamics, while an N-body code was used for the collisionless component consisting of dark matter and galaxies. They showed the development of outward propagating shock waves generated during  the collisions between subclusters. 

None of the works mentioned above included the baryonic contribution to the gravitational potential.
Head-on equal mass collisions between clusters were simulated by \cite{1994MNRAS.268..953P} using smoothed-particle hydrodynamics (SPH), including both baryons and dark matter. \cite{1997ApJS..109..307R} used a PPM/particle-mesh code to simulate a collision intended to resemble the cluster Abell~754. 
\cite{1999ApJ...518..594R} presented the first three-dimensional numerical magnetohydrodynamical (MHD) simulations of the magnetic field evolution in merging clusters of galaxies, which they subsequently applied to model the double radio relics in Abell~3367 \citep{1999ApJ...518..603R}. They concluded that the radio structures arise from a slightly off-axis merger that occurred nearly in the plane of the sky approximately 1~Gyr ago, with the subcluster having a total mass of 20\% of the primary cluster.
\cite{2000ApJ...538...92R} also modeled the X-ray substructure seen in Abell~3266 as an off-axis merger with a mass ratio of $2.5:1$.

\cite{1998ApJ...496..670R} simulated off-center collisions of equal mass clusters using a PPM code and  an isolated multigrid potential solver. They simulated off-center collisions with 0, 5, and 10 times the cluster core radius and studied how the virialisation time, X-ray luminosity, and structure of the merger depended on the impact parameter of the collision. They found that the increase in X-ray luminosity due to the merger depends on the impact parameter of the collision. \cite{2002MNRAS.329..675R} also reported that mergers can lead to large increases in bolometric X-ray luminosities and emission-weighted temperatures. Cool cores are completely disrupted by equal mass head-on mergers, while for mergers with mass ratios of $8:1$ the cooling flow restarts within a few Gyr. For mergers with impact parameters of 500~kpc the cool-core is not disrupted. \cite{2001ApJ...561..621R}  investigated cluster collisions, varying both the mass ratio and impact parameters. They found that merger events created large-scale turbulent motions with eddy sizes up to several hundred kiloparsecs.

FLASH \citep{2000ApJS..131..273F} high-resolution N-body/hydrodynamics simulations  have been carried out by \cite{2009ApJ...699.1004Z} to model the cluster Cl~0024+17, a proposed merger of two clusters, with the interaction occurring along our line of sight. Also, gas sloshing initiated by mergers with subclusters, and the effects of mergers on the entropy of the ICM  were investigated by \cite{2010ApJ...717..908Z, 2011ApJ...728...54Z}. Various simulations of cluster merger were also presented by \cite{1999ApJ...520..514T, 2000ApJ...532..183T, 2000ApJ...535..586T, 2008ApJ...687..951T}. In \cite{2008ApJ...687..951T} N-body + magnetohydrodynamical simulations were performed from which is was found that cluster mergers cause various characteristic magnetic field structures. The magnetic field component perpendicular to the collision axis is amplified by the merger which results in  a cool region wrapped by field lines. \cite{2007MNRAS.376..497M} focussed on the generation of entropy during cluster mergers and \cite{2010PASJ...62..335A} investigated how the ionization equilibrium state is affected by cluster mergers. They report that the ICM significantly departs from the ionization equilibrium state at the location of shocks.

\cite{2006MNRAS.373..881P} carried out a suite of SPH simulations of merger of clusters  whose initial conditions resembled relaxed cool core clusters. They investigated 
the efficacy of (i) centroid variance, (ii) power ratios, and (iii) X-ray surface brightness/projected mass displacement  to quantify the degree to which observed clusters are disturbed. From this it was found that the centroid variance gives the best measure of the state of clusters and how far they are from virial and hydrostatic equilibrium.

Cluster mergers can decouple the baryonic matter component from the dark matter (DM) which causes  and offset between the gravitational center (measured from lensing)  and X-ray center of the cluster. This was first observed for the ``Bullet cluster''  \citep[1E0657$-$56,][]{2006ApJ...648L.109C}. \cite{2007MNRAS.380..911S} presented hydrodynamical models of galaxy cluster mergers to reproduce the dynamical state and mass models (from gravitational lensing) of the ``Bullet'' cluster (1E0657$-$56). They showed that the size of the spatial offset between the baryonic peak and mass is quite sensitive to the structural details of the merging systems. This offset was earlier also found in numerical models by \cite{2006PASJ...58..925T}. \cite{2008MNRAS.389..967M} presented detailed  N-body/SPH simulations of the system. They found that the  X-ray morphology is best simulated with a mass ratio of $6:1$ and an impact parameter of  150~kpc.

\section{Numerical method}
\label{sec:method}
We used the FLASH 3.2 framework \citep{2000ApJS..131..273F} for the simulations which includes standard hydrodynamics and gravity. We did not include the effects of radiative cooling, as this is not very relevant for the spatial and temporal time scales considered here. 
Furthermore, the hydrogen number density was assumed to be related to the electron number density as $n_{\rm{H}} = 0.6n_{\rm{e}}$. 
We chose outflow boundary conditions. We simulated a box with a size of $5 \times 5 \times 5$~Mpc. A maximum of 6 refinement levels are allowed, resulting in a maximum resolution of 3.25~kpc. The minimum refinement level was set to 3 giving a resolution of 26~kpc. The simulations were run at this relatively low-resolution in order to study a larger range of merger parameters.

For the gravitational potential of each subcluster we assume hydrostatic equilibrium and spherical symmetry, 
\begin{equation}
\frac{d\phi}{dr} = \frac{k_{\rm{B}} }{\mu m_{\rm{H}}}\left[\frac{T(r)}{\rho(r)}\frac{d{\rho(r)}}{d\log{r}} + \frac{d{T(r)}}{d\log{r}}     \right] \mbox{ ,}
\label{eq:mr}
\end{equation}
where $T(r)$ and $\rho(r)$ are the radial density and temperature profiles, $m_{\rm{H}}$ the hydrogen mass,  and $\mu$ the mean molecular weight (we adopt $\mu=0.6$). We move the center of the gravitational potential of the merging subcluster around the fixed potential of the main cluster, ignoring the interactions between the dark matter.

\subsection{Radio emission from shocks}
\label{sec:radiomodel}

We use passive tracer particles in the simulations to model the radio emission. At the start of the simulation the particles are distributed proportional to the density throughout the complete computation volume. The advantage is that in this way each passive tracer particle represent roughly the same amount of mass which simplifies the normalization of the radio emission, as the contribution from every particle can be simply added up. For each tracer particle, 5 million in total, we record the position, velocity, density and temperature as a function of time, writing the results to disk every $30 \times 10^{6}$~yr. In order to create synthetic radio maps, we locate all particles that have passed through the shock waves in the simulation by looking at the variation of the temperature as function of time for each tracer particle. We define a particle to have passed through a shock if the temperature between two successive outputs spaced $30 \times 10^{6}$~yr apart increases by more than a factor of 1.5 (no radio emission is generated below this number). For each particle that passes through the shock we then compute the Mach number ($\mathcal{M}$, using Rankine-Hugoniot jump conditions and adiabatic index $\gamma =5/3$ \citep{1959flme.book.....L}, compression ratio, entropy ratio, and the time since it has passed through the shock. With these parameters we  compute the synchrotron emission at a given frequency $\nu$ using the JP model \citep{1973A&A....26..423J}, taking into account the spectral ageing due to synchrotron and IC losses. We apply eq. 5 and 6 from \cite{1994A&A...285...27K} and use $t_0 = 0$, see also \cite{1962SvA.....6..317K, 1970ranp.book.....P}.

The amount of synchrotron emission for each tracer particle is recorded.  A radio map is then simply computed by integrating the radio emission from each tracer particle in the computational volume along a chosen line of sight. 

We take $B_{\mathrm{CMB}}=4.6$~$\mu$Gauss the equivalent magnetic field strength of the CMB at $z = 0.1921$. To compute the radio emission the magnetic field is needed. Since we do not include magnetic fields in our simulation we adopt constant values for the magnetic field $B$. 
The spectra are then normalized using the method described by \cite{2007MNRAS.375...77H}, which takes into account the efficiency of acceleration as function of Mach number. We start with a spectrum of suprathermal electrons with a power-law distribution $n(E) \propto E^{-s}$, with $E$ the energy. The index $s$, with $s=1-2\alpha_{\rm{inj}}$, and $\alpha_{\rm{inj}}$ the injection radio spectral index, depends only on the 
compression ratio (or Mach number), $r$, of the shock front:
\begin{equation}
s = \frac{r+2}{r-1}
\end{equation}
and \citep[e.g.,][]{2008A&A...486..347G}
\begin{equation}
\alpha_{\rm{inj}} =  \frac{1}{2} - \frac{\mathcal{M}^2 +1} {\mathcal{M}^2 -1} \mbox{ .}
\label{eq:inj-mach}
\end{equation}

We assume that a small fraction $\xi_{\rm{e}}$ of the thermal energy goes into the acceleration of electrons \citep{2004ApJ...617..281K}. Furthermore, we take into account that the electrons can only be accelerated to a finite maximum energy ($E_{\rm{max}}$) \citep{2003ApJ...585..128K}. The electron spectrum with a smoothed high-energy cut-off is then given by
{\small
\begin{eqnarray}
  \frac{{\rm d}n_{\rm e}}
       {{\rm d}E}
  =
  \left\{
  \begin{array}{r@{\quad:\quad}l}
    n_{\rm e} \,
    C_{\rm spec} \:
    \frac{ 1 }{ m_{\rm e} c^2} \,
    \tilde{e}^{-s}
    \left\{
      1 
      - 
     \frac{ \tilde{e} }
          { \tilde{e}_{\rm max} }
    \right\}^{s-2}
    &
    \tilde{e} < \tilde{e}_{\rm max}
    \\
    0
    &
    {\rm elsewhere}
    \\
  \end{array}
  \right.
  ,
  \label{eq-e-supr-spec}  
\end{eqnarray}}where $\tilde{e}=E/m_{\rm{e}}c^2$. The normalization constant $C_{\rm{spec}}$ gives the fraction of electrons at $E=m_{\rm{e}}c^2$.

Following \cite{2007MNRAS.375...77H}, we assume that the thermal Maxwell-Boltzmann distribution goes
over continuously into the power-law spectrum of the suprathermal electrons. \cite{2007MNRAS.375...77H} show that the normalization of the electron spectrum is given by 
\begin{eqnarray} 
  C_{\rm spec}
  =
 {
    \xi_e
    \frac{u\down}{c^2}
    \frac{m_{\rm p}}{m_{\rm e}}}_{} \;
 {
    \frac{(q-1)}{q}
    \frac{1}{I_{\rm spec}}
  }_{}
  ,
  \label{eq-C1}
\end{eqnarray}
where ${u\down}$ is the downstream velocity, $q$ the entropy ratio ($S\down/S\up$) across the shock, and
\begin{eqnarray} 
I_{\rm spec} = 
\int_{\tilde{e}_{\rm min}}^\infty {\rm d}\tilde{e} \;
  \tilde{e}^{1-s} \,
  \left\{
   1 
   -
   \frac{\tilde{e}}
        {\tilde{e}_{\rm max}}
   \right\}^{s-2}
   .
\end{eqnarray}
The minimum energy, $\tilde{e}_{\rm{min}}=E_{\rm min}/m_{\rm{e}}c^2$, is the energy above which electrons are considered to be suprathermal. This results is an implicit equation for $E_{\rm min}$ that has to be solved simultaneously with the normalization of the spectrum. We directly apply the code from \citeauthor{2007MNRAS.375...77H} for this.

\subsection{Initial conditions}
For the initial conditions we start with two spherically symmetric (sub)clusters with masses $M_1$ and $M_2$ separated by a distance $d$. Subscript $1$ always refers to the more massive cluster (i.e., $M_1 > M_2$). The density profiles of both subclusters are described by single $\beta$-models \citep{1976A&A....49..137C}

\begin{equation}
\rho(r) = \rho_0 \left[ 1+ \left(\frac{r}{r_{\rm{c}}}\right)^{2}    \right]^{-\frac{3\beta}{2}} \mbox{ ,}
\label{eq:betamodel}
\end{equation}
where $\rho_{0}$ the central density, $r_{\rm{c}}$ the core radius. We decided to fix $\beta$ to $2/3$ \citep[e.g.,][]{2000ApJS..129..435B} and $\rho_{0}=2 \times 10^{-26}$~g~cm$^{-3}$  (0.02 particles per cm$^{3}$) for both subclusters. This value for $\rho_{0}$ is roughly the average of the values found by \cite{2008A&A...487..431C} for a sample of 31 clusters. They found that the  central densities did not show any clear trend with the overall mass and temperature of the clusters.

We used the $M-T$ scaling relation from \cite{2010MNRAS.406.1773M} to get the global average temperature from the cluster's mass. 
\begin{equation}
T_{\rm{avg}}=  10^{0.88 + 0.48\log_{10}{(M/10^{14}\rm{M_{\odot} )}}}  \mbox{ [keV] } \mbox{ .}
\label{eq:tavg}
\end{equation}
For the radial temperature distribution we use a polytropic profile \cite[e.g.,][]{2002A&A...394..375P} 
\begin{equation}
T(r) = T_{\rm{avg}} \tau  \rho(r)^{\gamma_{\rm{T}}-1}
\end{equation}
with $\tau=1.1$ and $\gamma_{\rm{T}} = 1.2$ and $T_{\rm{avg}}$ given by Eq.~\ref{eq:tavg}.
In addition we allow for a temperature drop in the center of the cluster to simulate a relaxed cool core cluster \citep[e.g.,][]{1991A&ARv...2..191F,  2006PhR...427....1P}. The temperature decline is the central region is given by \citep{2001MNRAS.328L..37A, 2006ApJ...640..691V}
\begin{equation}
T_{\rm{cool}}(r) = \frac{ x +   T_{\rm{min}}/T_{\rm{avg}}}   {x+1} \mbox{ }, \mbox{ } x = \left(\frac{r}{r_{\rm{cool}}}\right)^{a_{\rm{cool}}} \mbox{ ,}
\label{eq:tcool}
\end{equation}
where $T_{\rm{min}}/T_{\rm{avg}}$ is the relative temperature drop in the center, ${r_{\rm{cool}}}$ the radius of the cool core, and $a_{\rm{cool}}$ controls de rate of the temperature decrease. The final temperature model is given by 
\begin{equation}
T_{\rm{full}}(r) = T(r)T_{\rm{cool}}(r) \mbox{ .}
\end{equation}

For the merger scenario we follow \cite{2002ASSL..272....1S}, where two subclusters merge at time $t_{\rm{merge}}$ (the age of the Universe at the time of the merger). The two clusters have a large initial separation, $d_0$. The value of $d_0$ does not strongly affect the merger because the infall velocity approaches free-fall from infinity. If we assume that the clusters dominate the mass in the region of the Universe they occupy 
\begin{equation}
d_0 \approx 4.5 \left(\frac{M_1+M_2}{10^{15}\rm{M_\odot}}\right)^{1/3}    \left(\frac{t_{\rm{merge}} } {10^{10} \mbox{ yr}}\right)^{2/3} \mbox{ [Mpc]} \mbox{ .}
\end{equation}
For CIZA~J2242.8+5301 ($z=0.1921$) $t_{\rm{merge}}$ is $11.0$~Gyr.

To speed up the simulations the centers of the two subclusters are placed at a distance $d=2$~Mpc from each other. This we define as $t=0$.
The relative velocity between the clusters at that starting point is given by \cite{2002ASSL..272....1S} as
{\footnotesize\begin{equation}
v \approx 2930 \left(\frac{M_1 + M_2}{10^{15}\rm{M_{\odot}}}  \frac{1-\frac{d}{d_0}}{1-\left(\frac{b}{d_0}\right)^{2}}   \right)^{1/2}    \left(\frac{d}{1 \mbox{ Mpc}}\right)^{-1/2}   \mbox{ [km s}^{-1}\mbox{],}
\label{eq:v}
\end{equation}}
where $b$ is the impact parameter of the collision. Using the starting location, initial velocities and masses of the subclusters we move the center of the gravitational potential of the merging subcluster ($M_2$) around the fixed potential of the ``main'' cluster ($M_1$), ignoring the interactions between the dark matter and treating the clusters as point masses. This fixes the subcluster's orbit around the main cluster. The point $(x,y,z) = (0,0,0)$ is defined by the center of mass of the system. The gravitational acceleration of the  gas was then computed using Eq.~\ref{eq:mr}. 

To reconstruct the merger that gave rise to the double radio relic CIZA~J2242.8+5301, we varied the mass ratio and impact parameter in the simulation. In the post-processing (to create the radio maps) we also varied the viewing angle. The total mass of the two (sub)clusters ($M_1+M_2$) was kept constant to $5.5 \times 10^{14}\rm{M_{\odot}}$. The values for $r_{\rm{c}}$ are fixed by the total mass of the system and the required mass ratio (these are reported in Table \ref{tab:runs}). In Sect.~\ref{sec:discussion} we will discuss the effect on the simulations if we take different values for $\beta$ and $r_{\rm{c}}$. For the default runs we chose $T_{\rm{min}}/T_{\rm{avg}}= 1$, i.e., no temperature drop in the cluster center. In Sect.~\ref{sec:coolcore} we discuss the effect of including a central cool core to the most massive subcluster. For all runs we place the merger axis along the y-axis with the least massive cluster located at positive y-value and the initial velocity (given by Eq.~\ref{eq:v} being in negative y-direction).

\section{Results}
\label{sec:results}

\begin{figure*}
\includegraphics[angle =90, trim =0cm 0cm 0cm 0cm,width=0.3\textwidth, clip=true]{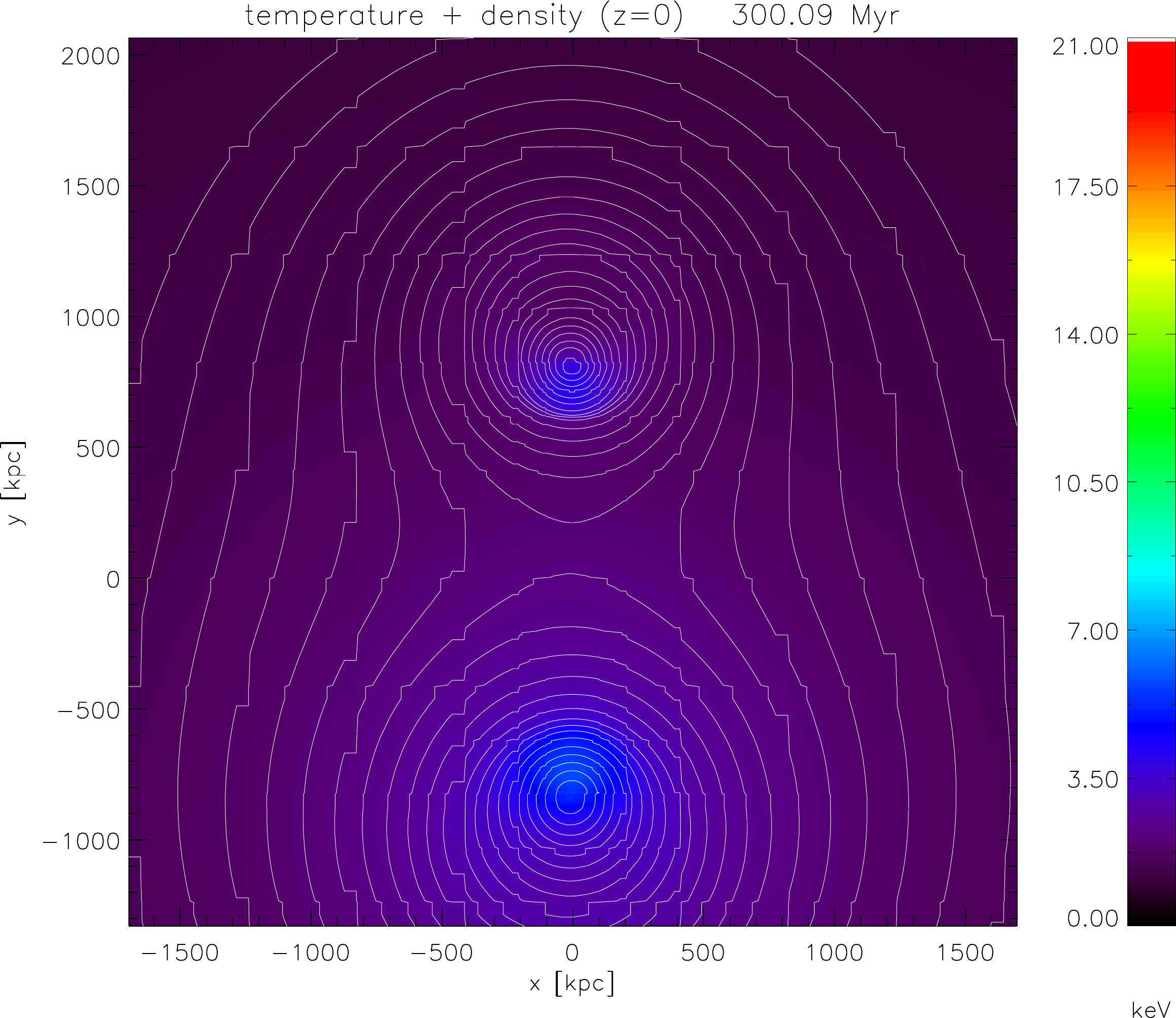}
\includegraphics[angle =90, trim =0cm 0cm 0cm 0cm,width=0.3\textwidth, clip=true]{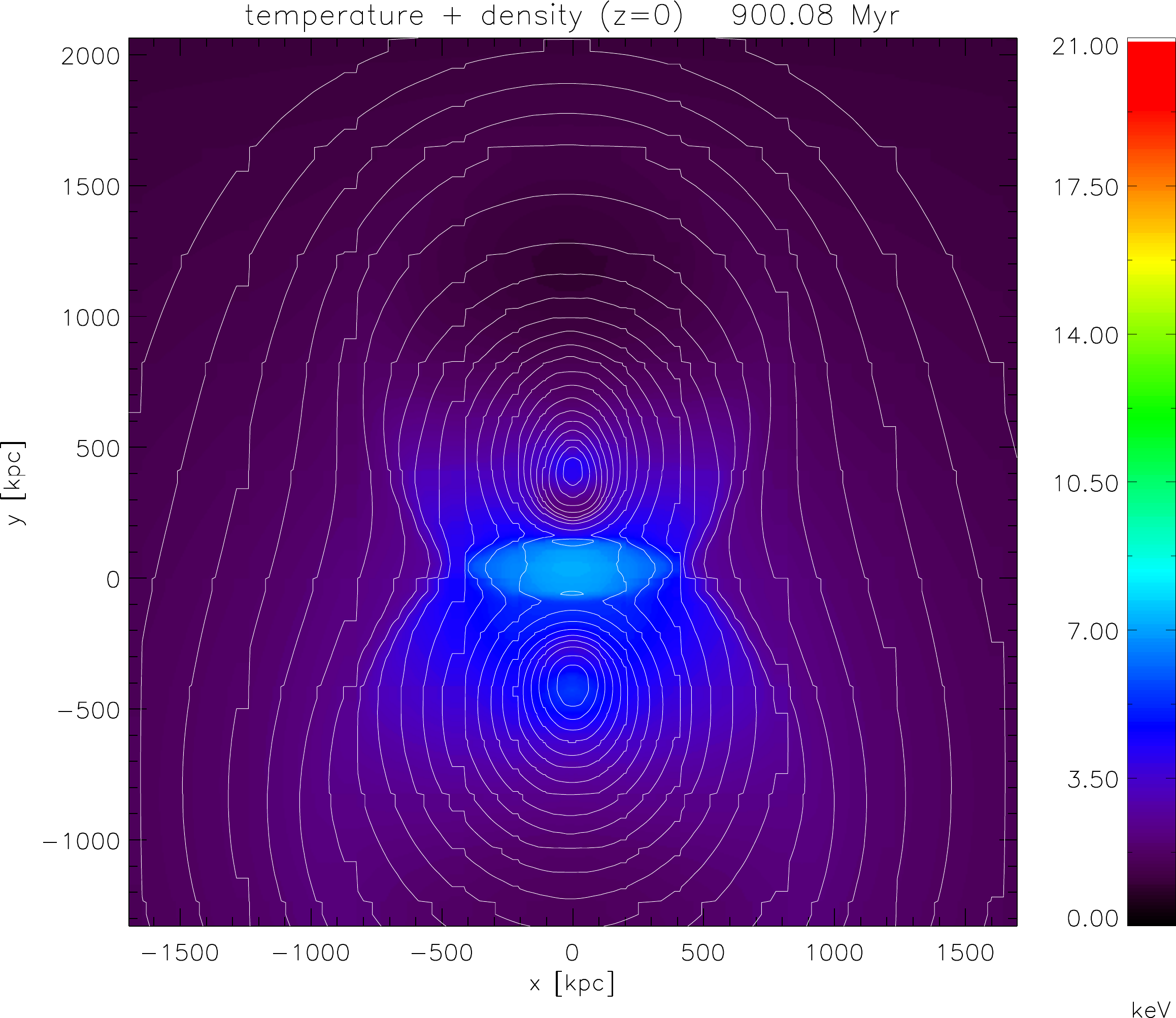}
\includegraphics[angle =90, trim =0cm 0cm 0cm 0cm,width=0.3\textwidth, clip=true]{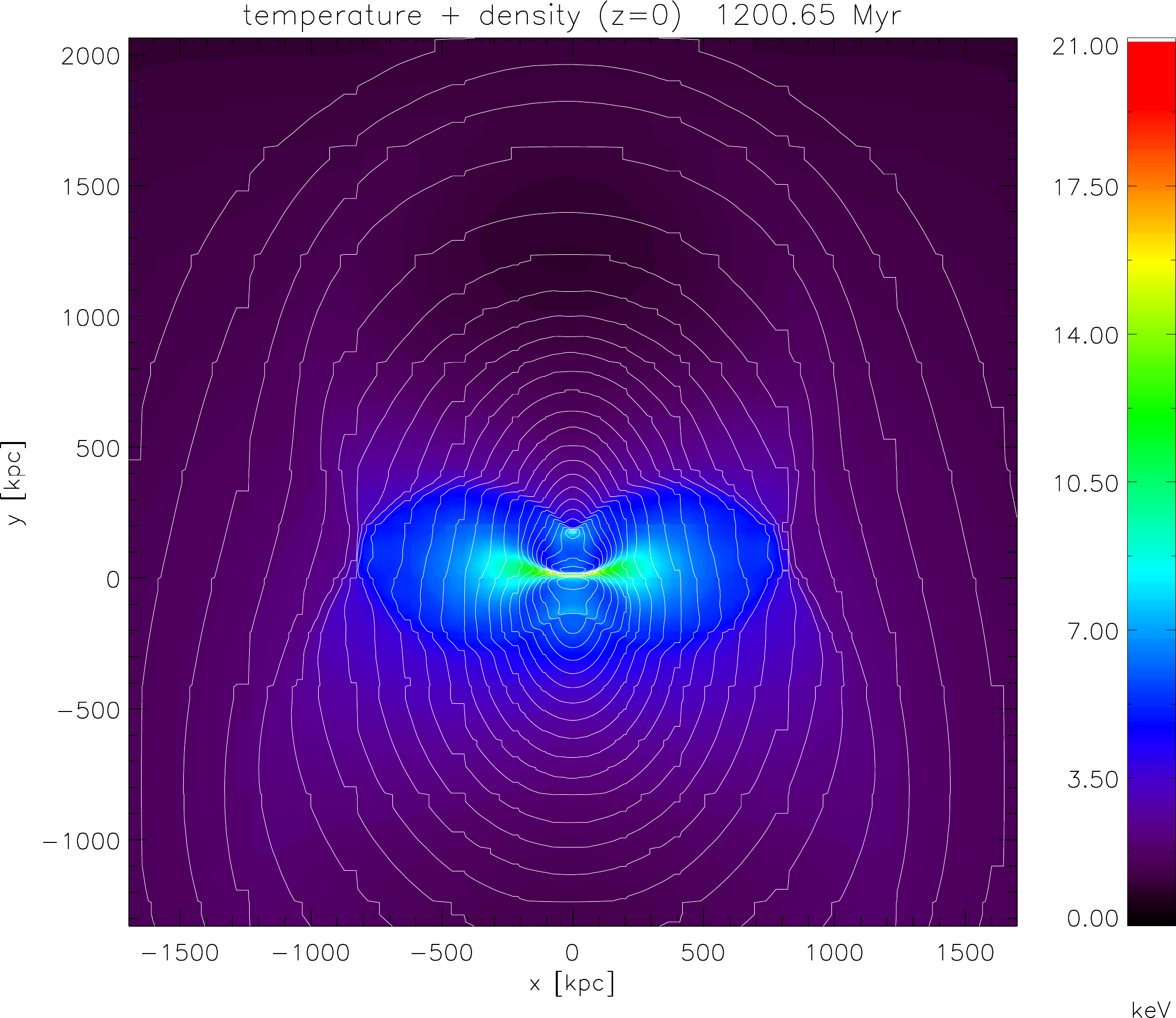}
\includegraphics[angle =90, trim =0cm 0cm 0cm 0cm,width=0.3\textwidth, clip=true]{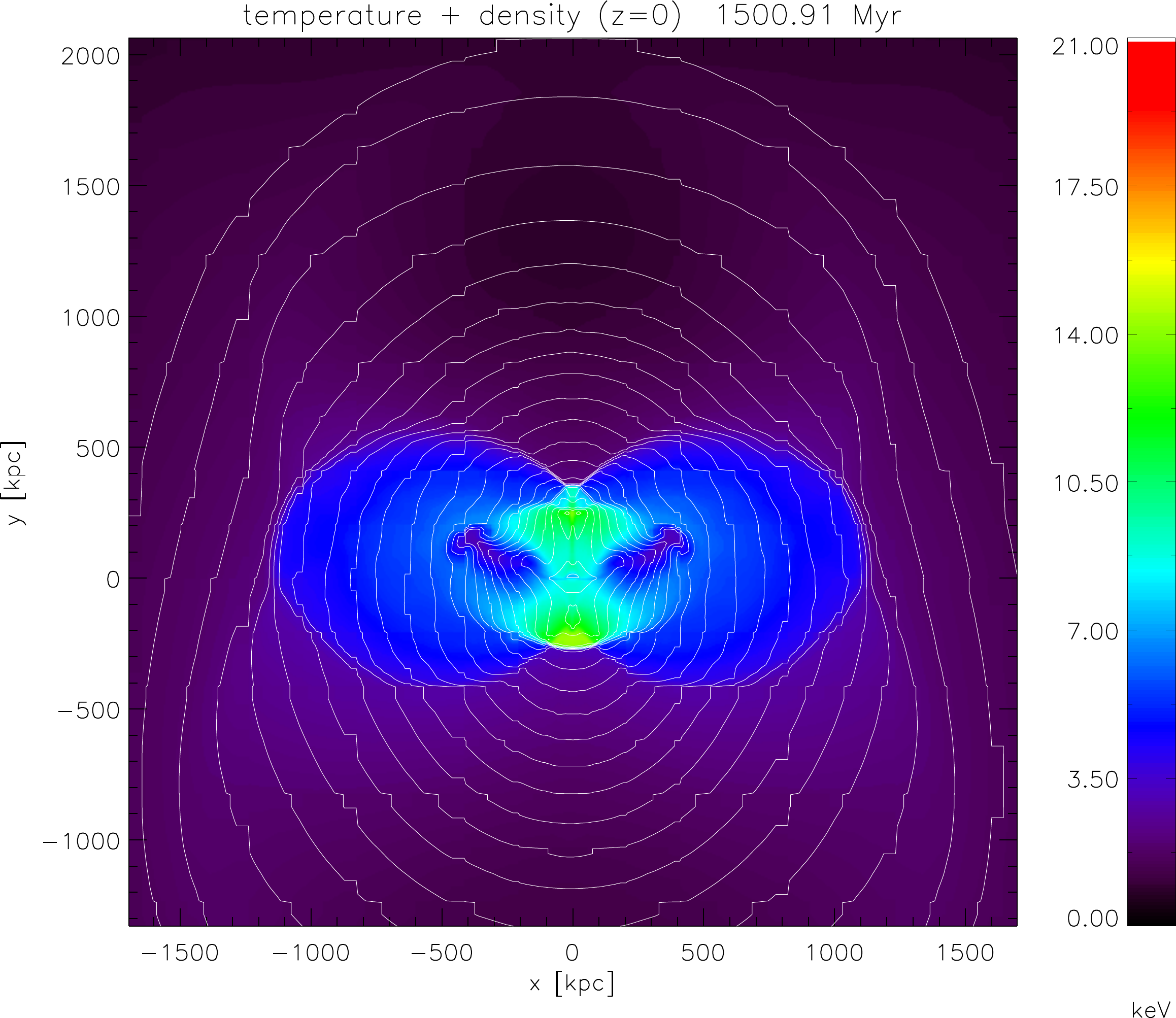}
\includegraphics[angle =90, trim =0cm 0cm 0cm 0cm,width=0.3\textwidth, clip=true]{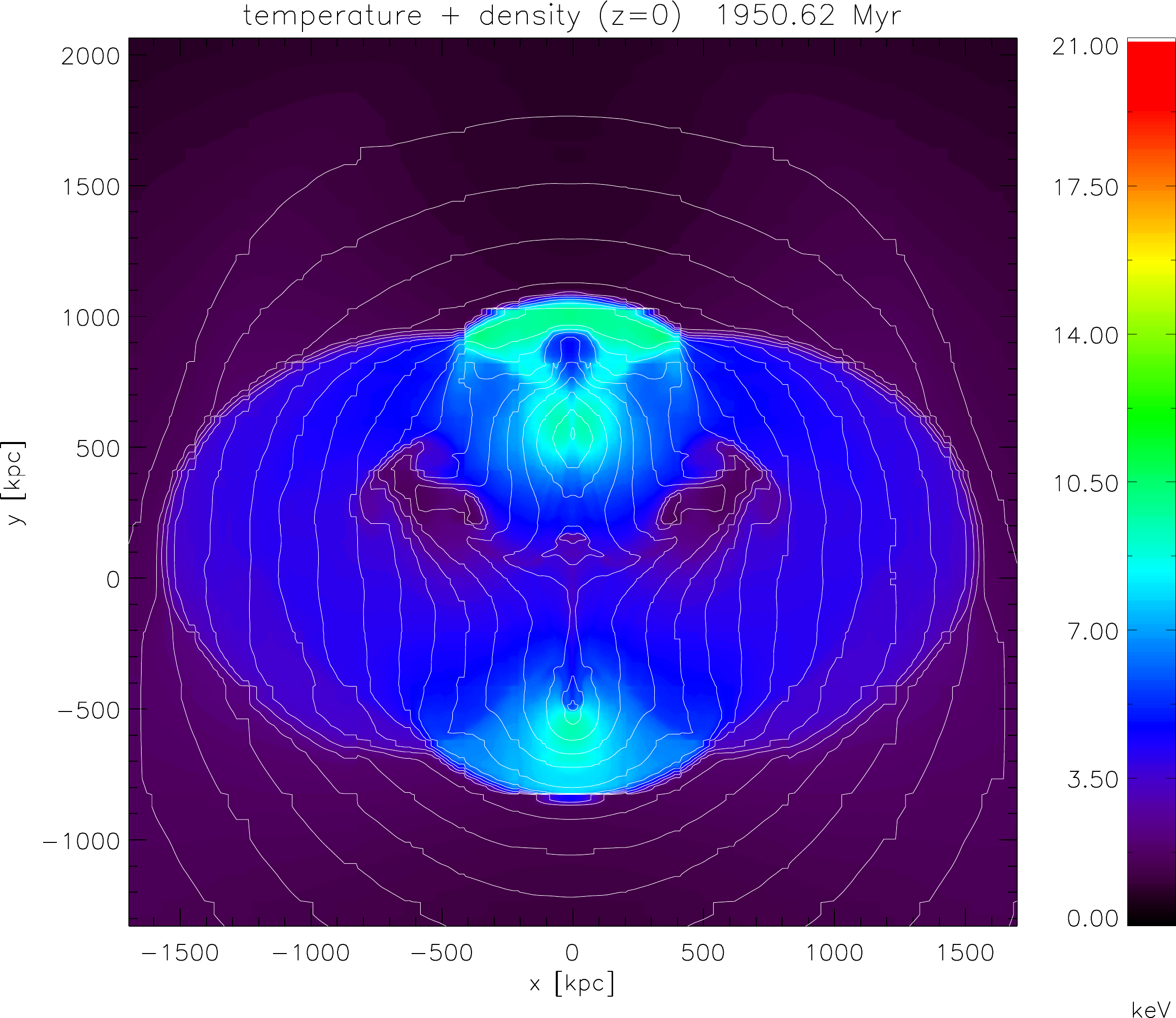}
\includegraphics[angle =90, trim =0cm 0cm 0cm 0cm,width=0.3\textwidth, clip=true]{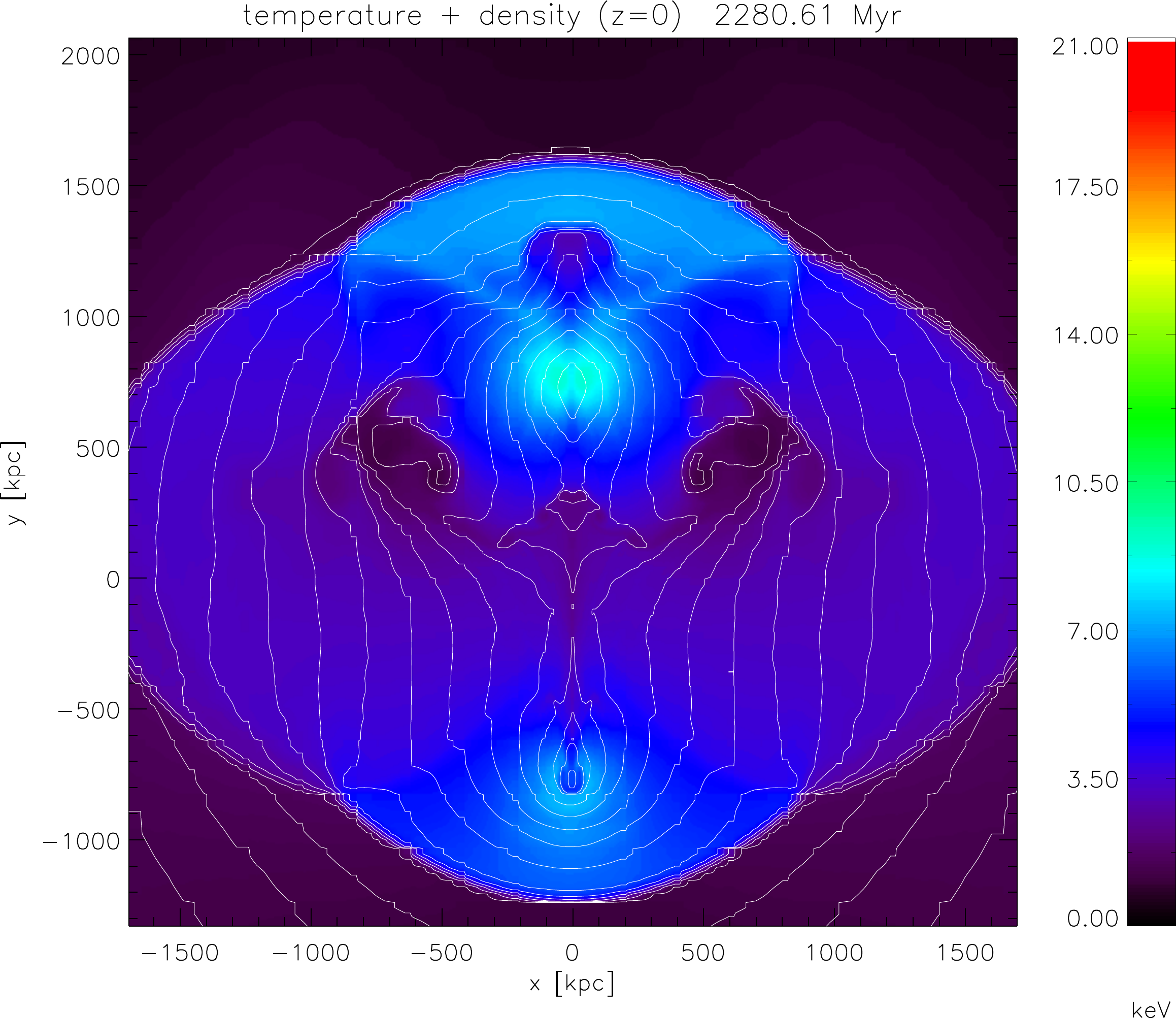}
\caption{Temperature and density evolution for a 2:1 merger with zero impact parameter. The color image displays the temperature at a slice $z=0$, while the white contours follow the density. The density contours are drawn at levels of $\log_{10}{(\rho (\rm[ g~cm^{-3}]}) = [-27.9,-27.8,-27.7,\ldots]$.} 
\label{fig:hydror21b0}
\end{figure*}

\begin{figure*}
\includegraphics[angle =90, trim =0cm 0cm 0cm 0cm,width=0.3\textwidth, clip=true]{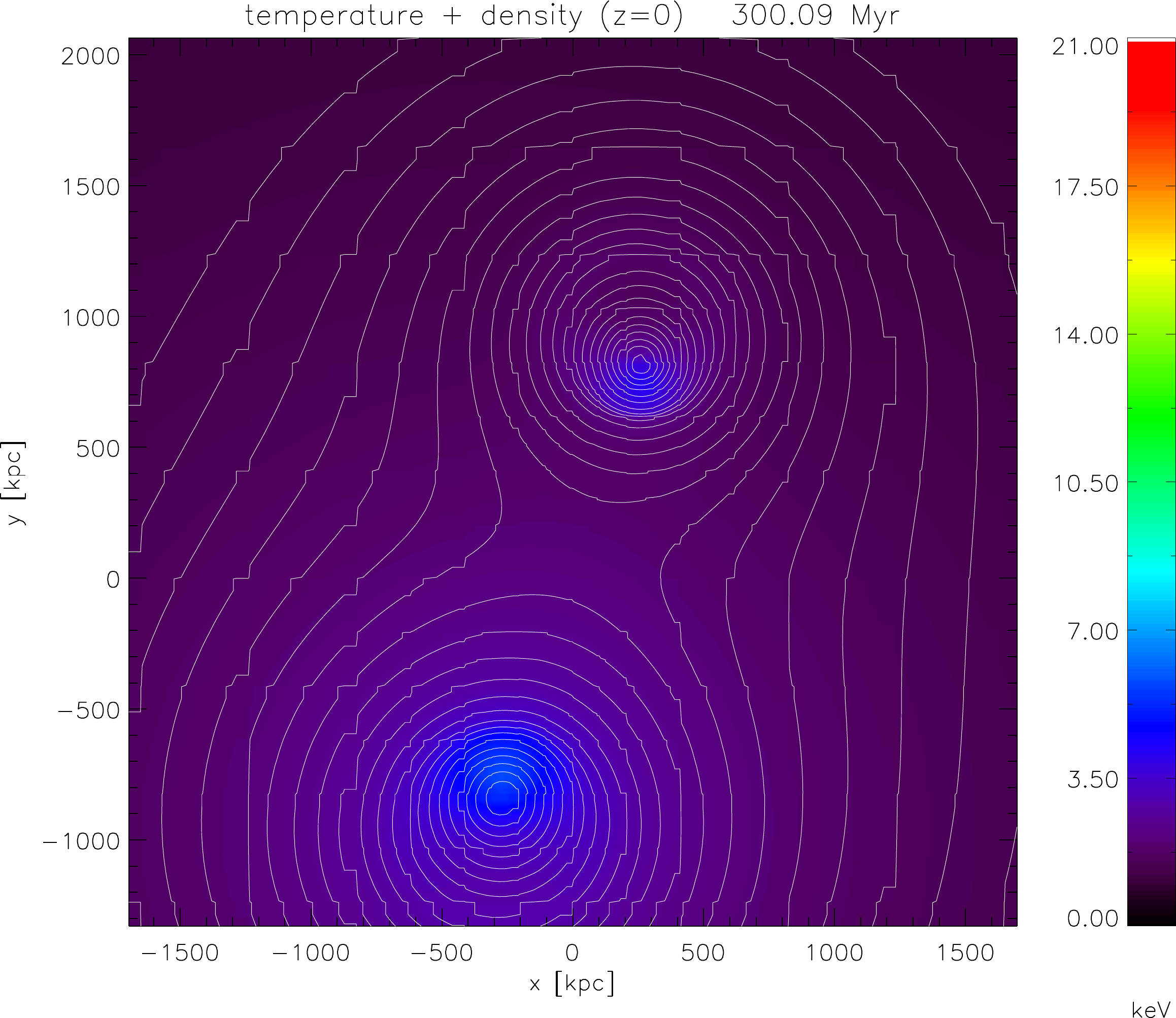}
\includegraphics[angle =90, trim =0cm 0cm 0cm 0cm,width=0.3\textwidth, clip=true]{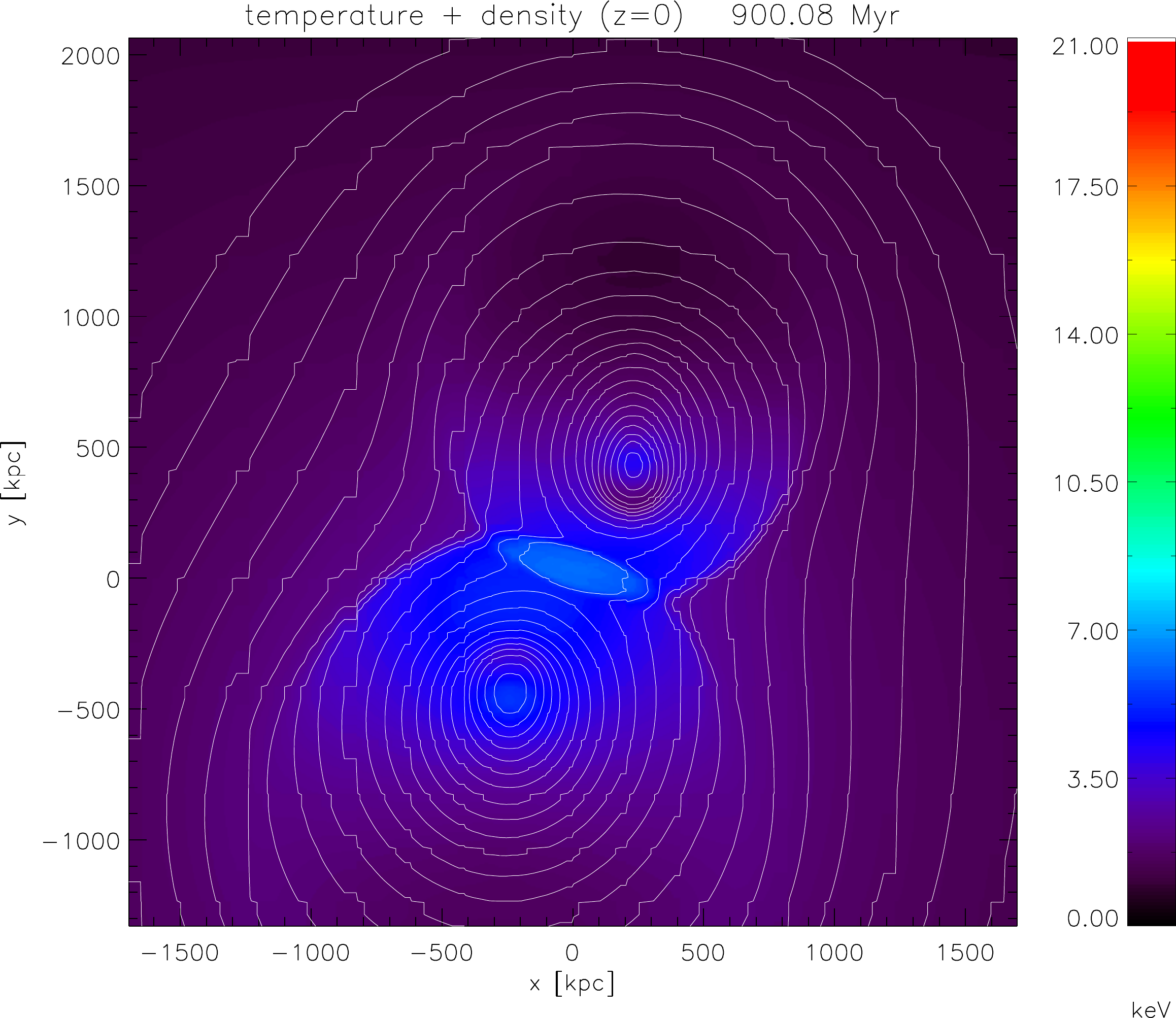}
\includegraphics[angle =90, trim =0cm 0cm 0cm 0cm,width=0.3\textwidth, clip=true]{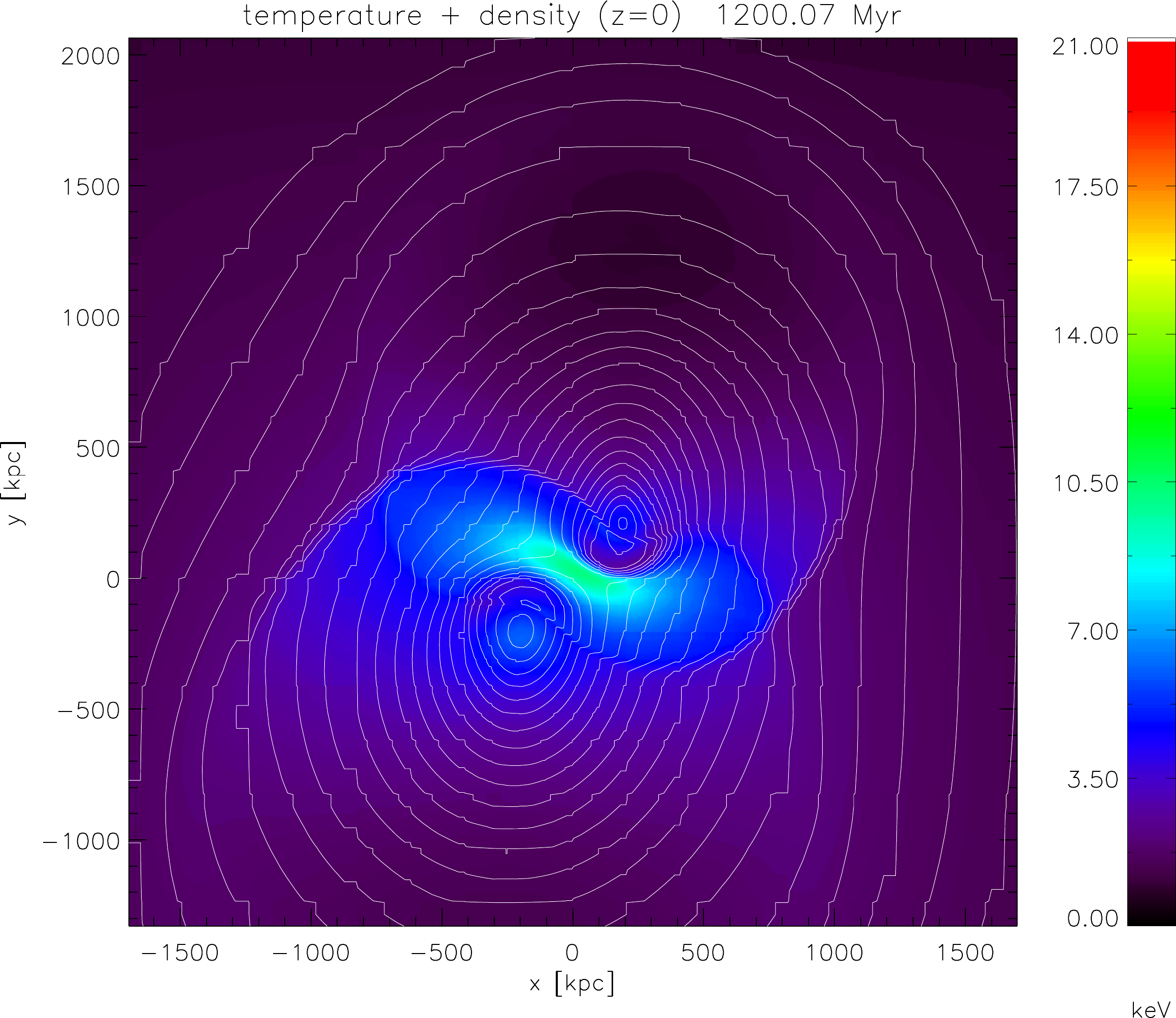}
\includegraphics[angle =90, trim =0cm 0cm 0cm 0cm,width=0.3\textwidth, clip=true]{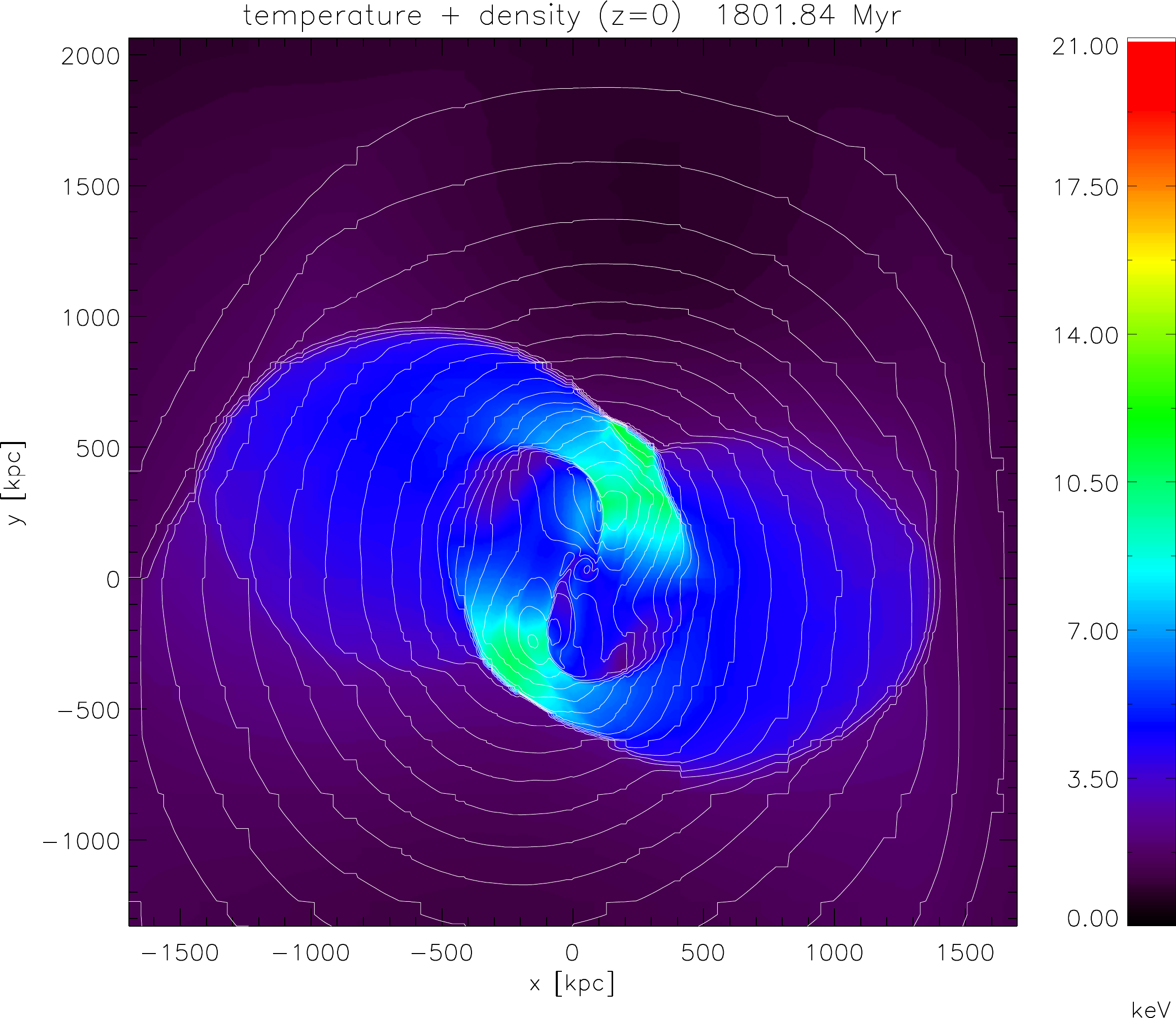}
\includegraphics[angle =90, trim =0cm 0cm 0cm 0cm,width=0.3\textwidth, clip=true]{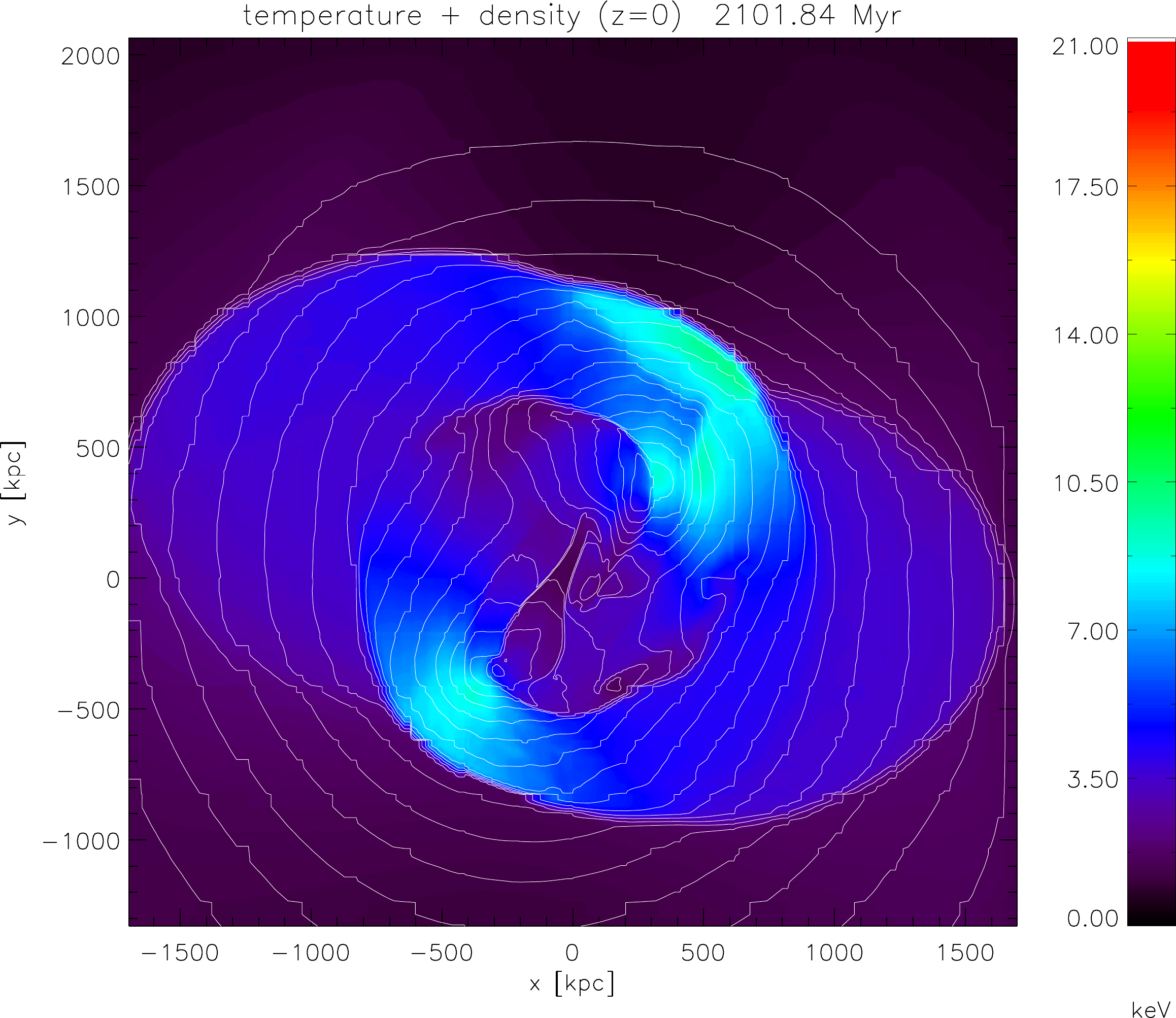}
\includegraphics[angle =90, trim =0cm 0cm 0cm 0cm,width=0.3\textwidth, clip=true]{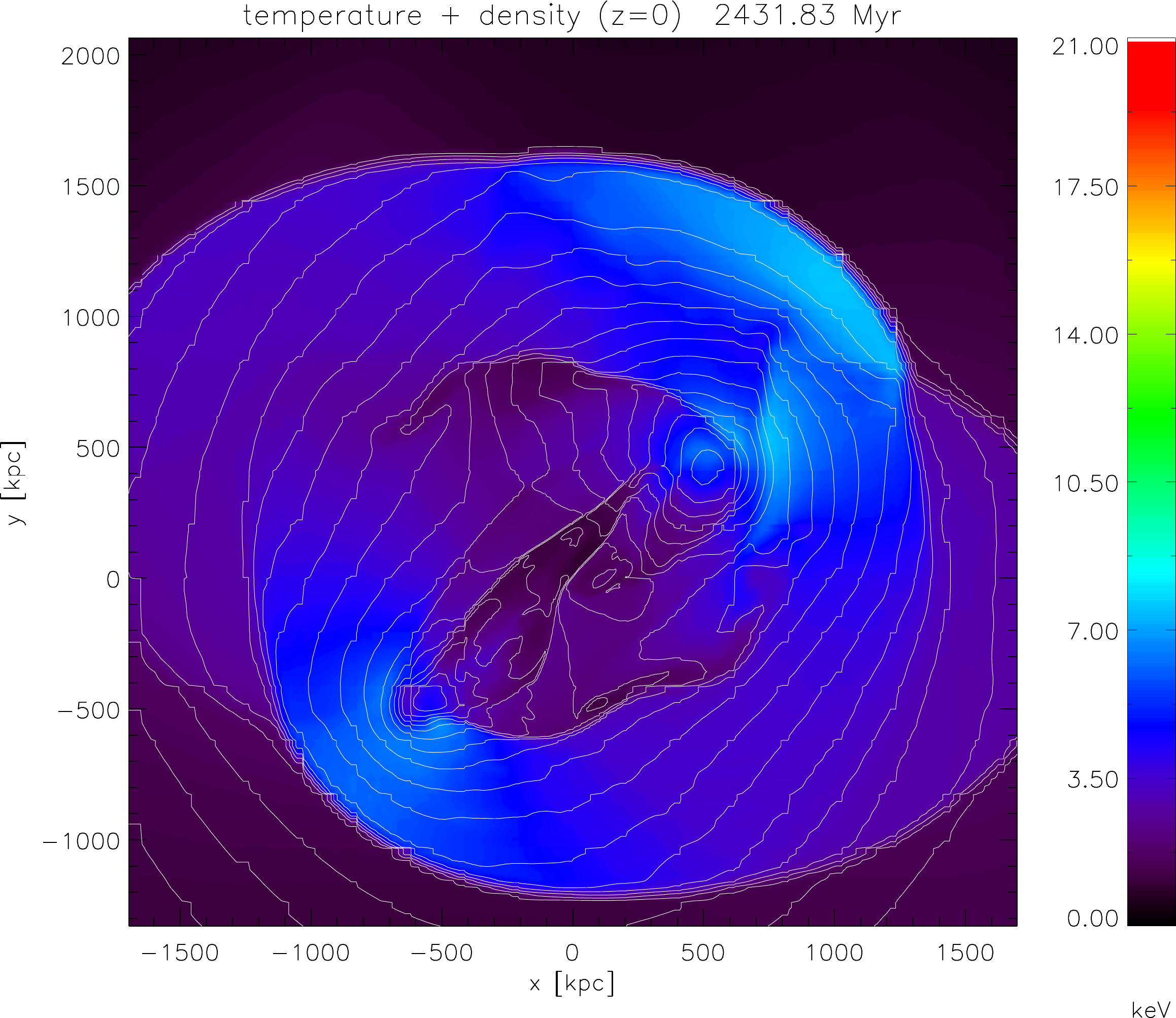}
\caption{Temperature and density evolution for a 2:1 merger with an impact parameter $b=4r_{\rm{c,1}}$. The color image displays the temperature at a slice $z=0$, while the white contours follow the density. The density contours are drawn at levels of $\log_{10}{(\rho (\rm[ g~cm^{-3}]}) = [-27.9,-27.8,-27.7,\ldots]$.} 
\label{fig:hydror21b4}
\end{figure*}

The names and parameters for the different FLASH runs are given in Table~\ref{tab:runs}. We start with a description of the basic hydrodynamical quantities (the density, temperature, and the velocity field) from the simulations. We will discuss the results for a merger with a mass ratio of $2:1$ and zero impact parameter (R21b0; see Fig.~\ref{fig:hydror21b0}), and for a merger with the same mass ratio but with an impact parameter of 4 ($4r_{\rm{c,1}}$; R21b4), see Fig.~\ref{fig:hydror21b4}. 

The R21b0 merger starts with the subcluster approaching the main cluster.  Two shocks, with $\mathcal{M} \sim 1.15$ at $t=600$~Myr, develop as the cores approach each other. After core passage, at $t=1.25$~Gyr, a pair of stronger symmetrical shock waves form that travel diametrically outwards in front of the two subclusters.  The shock waves increase in size as the time advances, although their Mach numbers decrease. At $t=1.95$~Gyr the maximum Mach numbers are about 3.1 and 2.3 for the top and bottom shock waves, respectively. At $t=2.28$~Gyr the two shock are separated by 2.8~Mpc, as is the case for the double relics in CIZA~J2242.8+5301. The Mach numbers at this time have decreased to about $2.7$ and $2.2$, for the top and bottom relics, respectively. Thus, the most massive cluster with the largest core radius develops the largest (and slightly stronger) shock wave in front of it after core passage. The velocity field at  $t=2.3$~Gyr  is shown in Fig.~\ref{fig:vfield} (left panel).

Mergers with different mass ratios, ranging between 1 to 5, show a similar picture. The size and strength of the shock waves depend on the mass ratio (which in turn determines  the core radii of the subclusters). In collisions with a larger mass ratio the shock waves in front of the less massive subcluster reduces in relative size and strength. 

A $2:1$ merger with an impact parameter of $b=4r_{\rm{c,1}}$ (R21b4) develops two asymmetric spiral-like shock waves, and again the two shocks have different sizes depending on the mass ratio used. The amount of asymmetry deepens on the impact parameter. The Mach numbers across these asymmetric shocks decrease more slowly on the left side of the merger axis. At $t=2.1$~Gyr the maximum Mach number is about 2.9 and 2.2 for the top and bottom shock waves, respectively. This decreases to about 2.7 and 2.1 at $t=2.4$~Gyr when the two shocks are separated by 2.8~Mpc. The velocity field at $t=2.4$~Gyr is shown in Fig.~\ref{fig:vfield} (right panel).

\subsection{Radio maps}
\label{sec:radiomap}
In this section we present the simulated radio maps for a range of mass ratios, impact parameters, viewing angles and radial density/temperature profiles, and compare them to the observed WSRT radio map at 1382~MHz. To create the radio maps, we integrated the radio emission from each tracer particle that passed through the shock along the line of sight. All radio maps are normalized to the peak flux in the simulated map and are smoothed with a 2-dimensional Gaussian of $50\times40$~kpc to match the resolution of the WSRT image. This smoothing is also needed as the radio emission is computed using a discrete number of particles with insufficient sampling to create images at the resolution of the hydro-grid.  For making the radio maps, we assumed a constant magnetic field strength of $B=5.0$~$\mu$Gauss, see \cite{2010Sci...330..347V}. 

\begin{figure*}
\includegraphics[angle =90, trim =0cm 0cm 0cm 0cm,width=0.49\textwidth]{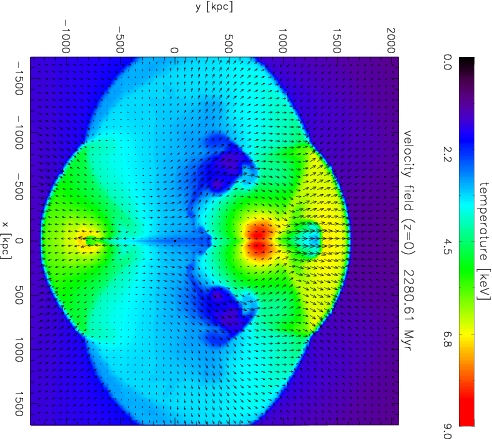}
\includegraphics[angle =90, trim =0cm 0cm 0cm 0cm,width=0.49\textwidth]{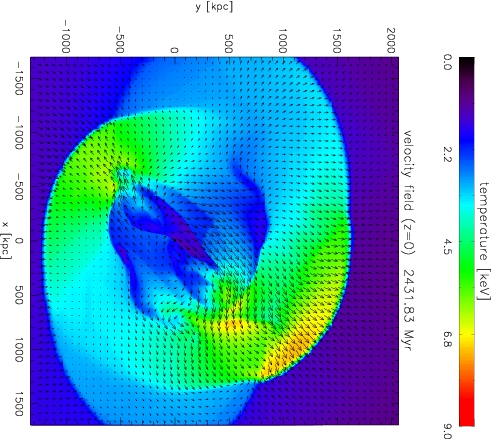}
\caption{Velocity field for a 2:1 merger with impact parameters $b=0$~kpc (left) and $b=4r_{\rm{c,1}}$ (right) for the slice $z=0$. The two snapshots were chosen such that the center of the two shock waves have a linear separation of 2.8~Mpc. The color image displays the temperature distribution. The maximum absolute velocity is $\sim1400$~km~s$^{-1}$.} 
\label{fig:vfield}
\end{figure*}

It takes about 1.0~Gyr for the two relics to be separated by 2.8~Mpc after core passage. This number varies very little for mass ratios between 1.5:1 and 3:1. For impact parameters $b> 0$~kpc the time since cores passage only increases slightly to about 1.1~Gr for $b=5r_{\rm{c,1}}$ (e.g., Fig.~\ref{fig:vfield}). Therefore, the time since core passage does not depend much on the adopted initial conditions such as the mass ratio and impact parameter. However, the time since core passage will also depend on the adopted total mass of the system and on the  temperature profile, since this affects the sound speed. Detailed X-ray observations would be helpful here to better constrain the total mass and measure the temperature profile.

\begin{table*}
\begin{center}
\caption{List of simulations with their defining parameters}
\begin{tabular}{lllllll}
\hline
\hline
run &   mass ratio$^{a}$ & impact parameter $(b)$ & $r_{\rm{c,1}}$, $r_{\rm{c,2}}$ & $T_{\rm{avg,1}}$, $T_{\rm{avg,2}}$ & $v_{\rm{y}}$$^{c}$  &comment\\
       &  $M_1:M_2$  & kpc& kpc, kpc & keV, keV & km~s$^{-1}$\\
\hline
R11b0     &  1:1       & 0 & 115, 115  & 4.1, 4.1 & $-937$  \\
R1.51b0  &  1.5:1   & 0 & 126, 101& 4.5, 3.7 & $-927$\\  
R21b0     &  1:1       & 0 & 134, 90 & 4.7, 3.3 & $-922$  \\
R2.51b0  &  2.5:1   & 0 & 139, 84 & 4.8, 3.1 & $-926$ \\ 
R31b0     &  3:1  & 0 & 145, 80 & 5.0, 3.0 & $-944$\\

R21b0.5  &  2:1   & $1r_{\rm{c,1}} =67$~kpc & 134, 90& 4.7, 3.3& $-922$\\  
R21b1  &  2:1   & $1r_{\rm{c,1}} = 134$~kpc & 134, 90& 4.7, 3.3& $-921$\\  
R21b2& 2:1 & $2r_{\rm{c,1}} = 268$~kpc      & 134, 90&4.7, 3.3 & $-916$\\ 
R21b3 &  2:1   & $3r_{\rm{c,1}} = 402$~kpc  & 134, 90&4.7, 3.3 & $-908$\\ 
R21b4 & 2:1 & $4r_{\rm{c,1}} = 536 $~kpc    & 134, 90&4.7, 3.3 & $-897$\\ 
R21b5 & 2:1 & $5r_{\rm{c,1}} = 670 $~kpc    & 134, 90&4.7, 3.3 & $-883$\\ 

R21b0$\beta0.5$ & 2:1 & 0 &82.6, 90   & 4.7, 3.3 & $-922$  & $\beta_1=0.5$ \\ 
R21b0$\beta0.75$ & 2:1 & 0 &162.5, 90 & 4.7, 3.3&  $-922$ &$\beta_1=0.75$ \\

R21b0cc & 2:1 & 0 &  134, 90 & 4.7, 3.3& $-922$ &cool core$^{b}$ for cluster 1\\
R21b0ss (A=0.2, $\lambda=100$~kpc)& 2:1 & 0 &  134, 90 & 4.7, 3.3& $-922$ &substructure\\ 

R21b0ss (A=0.3, $\lambda=75$~kpc)  & 2:1 & 0 &  134, 90 & 4.7, 3.3& $-922$ &substructure\\ 
R21b0ss (A=0.3, $\lambda=150$~kpc) & 2:1 & 0 &  134, 90 & 4.7, 3.3& $-922$ &substructure\\ 
R21b0ss (A=0.3, $\lambda=200$~kpc) & 2:1 & 0 &  134, 90 & 4.7, 3.3& $-922$ &substructure\\ 

R21b0ss (0.4, 200~kpc)& 2:1 & 0 &  134, 90 & 4.7, 3.3& $-922$ &substructure\\ 
\hline
\hline
\end{tabular}
\label{tab:runs}
\end{center}
$^{a}$ subscript 1 refers to the most massive subcluster, $M_1 + M_2 = 5.5\times 10^{14}$~M$_{\odot}$\\
$^{b}$ $T_{\rm{min,1}}/T_{\rm{avg, 1}} = 0.3$, $r_{\rm{cool,1}} = 50$~kpc and , $a_{\rm{cool,1}}=2$\\
$^{c}$ starting velocity in y-direction for the least massive cluster 2 (Eq.\ref{eq:v})\\
\end{table*}

\subsubsection{Mass ratio}
\label{sec:massratio}
\begin{figure*}
\includegraphics[angle =90, trim =0cm 0cm 0cm 0cm,width=1.0\textwidth, clip=true]{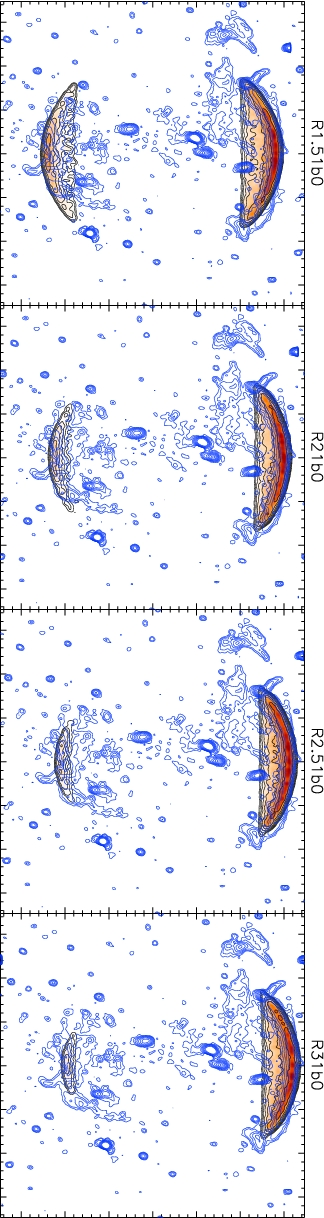}
\includegraphics[angle =90, trim =0cm 0cm 0cm 0cm,width=1.0\textwidth, clip=true]{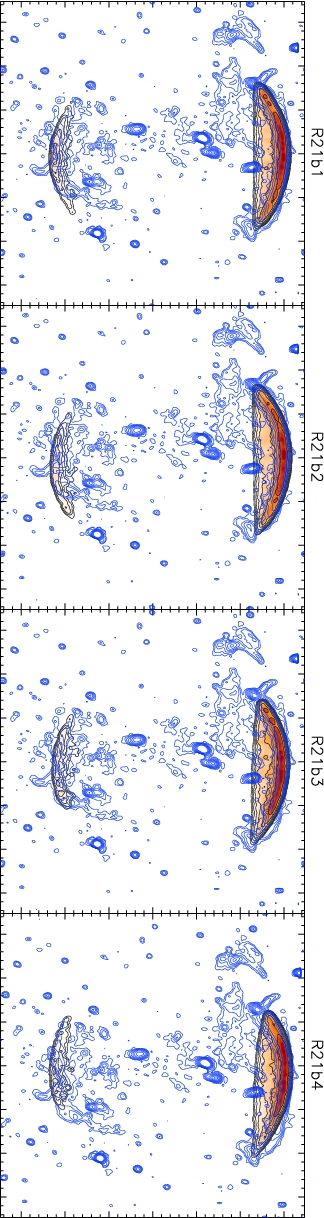}
\includegraphics[angle =90, trim =0cm 0cm 0cm 0cm,width=1.0\textwidth, clip=true]{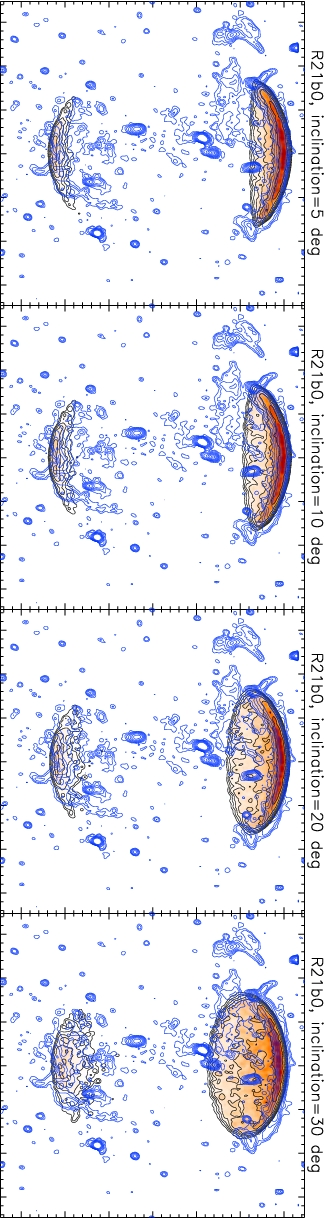}
\caption{Simulated radio emission for cluster mergers with different mass ratios (top row), impact parameters (middle row) and viewing angles (bottom row). The radio maps  to the hydrodynamical snapshot where the two shock are separated by 2.8~Mpc. The images span $3.4 \times 3.4$~Mpc. Blue contours show the observed emission at 1382~MHz from CIZA~J2242.8+5301 with the WSRT. The orange colored image and black contours display the synthetic radio image. Contours are spaced by factors of two in brightness}. 
\label{fig:massratio}
\end{figure*}

We varied the mass ratios for the collisions between $1:1$ and $3:1$.  
The mass ratio (which also set the core radii) affects the size of the two outwards moving shock waves after core passage. 
The resulting radio maps for mergers with four different mass ratios are displayed in Fig.~\ref{fig:massratio} and the largest linear extent of the relics are listed in Table~\ref{tab:ratios}. The radio maps display a double relic system, with the front of the two relics tracing the shock waves. Towards the cluster center and the left/right edges of the relics, the surface brightness decreases. The size of the shock waves (the $\mathcal{M} \gtrsim 2$ part) directly reflects the extent of the relics in the radio maps. 

We compare the synthetic radio maps to the observed WSRT map at 1382~MHz. 
The observed and synthetic maps are aligned by eye to the brightest northern relic in CIZA~J2242.8+5301, by simple translation and rotation. 
With a visual comparison and the taking into account the sizes listed in Table.~\ref{tab:ratios}, we find the best match between the observed and simulated radio maps for a mass ratio around $2:1$ to $1.5:1$ For more equal mass ratios the bottom relic increases to a size that is larger than the observed size. The opposite happens for larger mass ratios. When we compare the simulated radio images with the temperature map in Fig.~\ref{fig:vfield} (left panel) we note that the relic emission does not trace the full extent of the shock waves.  This is a consequence of the non-linear relation between the Mach number and normalization of the radio spectrum, discussed in Sect.~\ref{sec:radiomodel}. Below $\mathcal{M} \sim 2$ there is no observable radio emission generated. Another reason for the gradual disappearance of the radio emission at the right/left edges of the relics is that the amount of projected shock surface along the line of sight decreases.

\begin{table}
\begin{center}
\caption{Relic sizes}
\begin{tabular}{lllllll}
\hline
\hline
 &  top relic & bottom relic \\
\hline
CIZA~J2242.8+5301  &1700 &1450 \\
run R1.51b0 & 1730 &1570\\
run R21b0    & 1670 & 1220\\
run R2.51b0 & 1680 & 930\\
run R31b0    & 1690 & 660\\
\hline
\hline
\end{tabular}
\label{tab:ratios}
\end{center}
\end{table}

\subsubsection{Impact parameter}
\begin{figure}
\includegraphics[angle =90, trim =0cm 0cm 0cm 0cm,width=0.49\textwidth, clip=true]{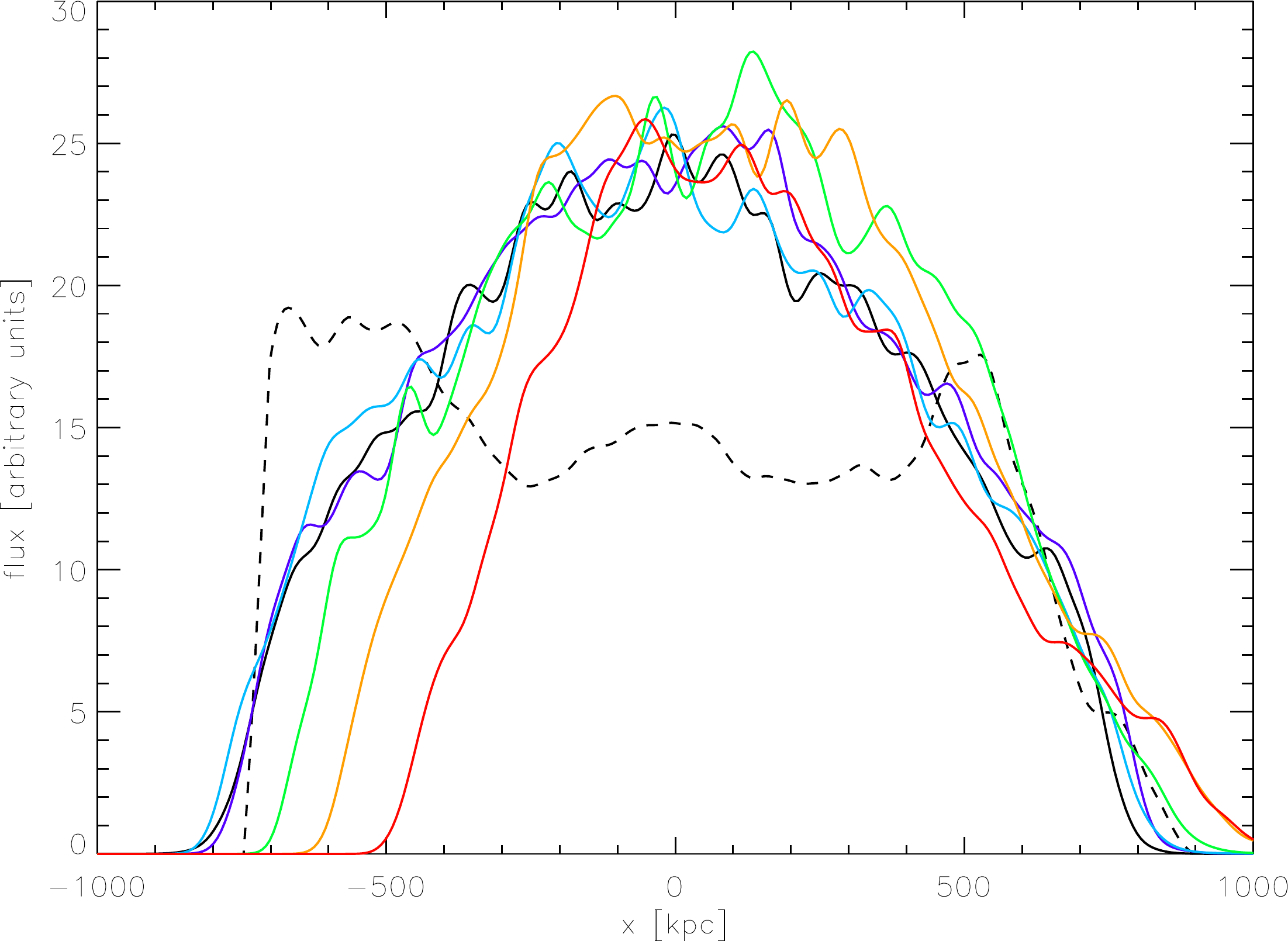}
\includegraphics[angle =90, trim =0cm 0cm 0cm 0cm,width=0.49\textwidth, clip=true]{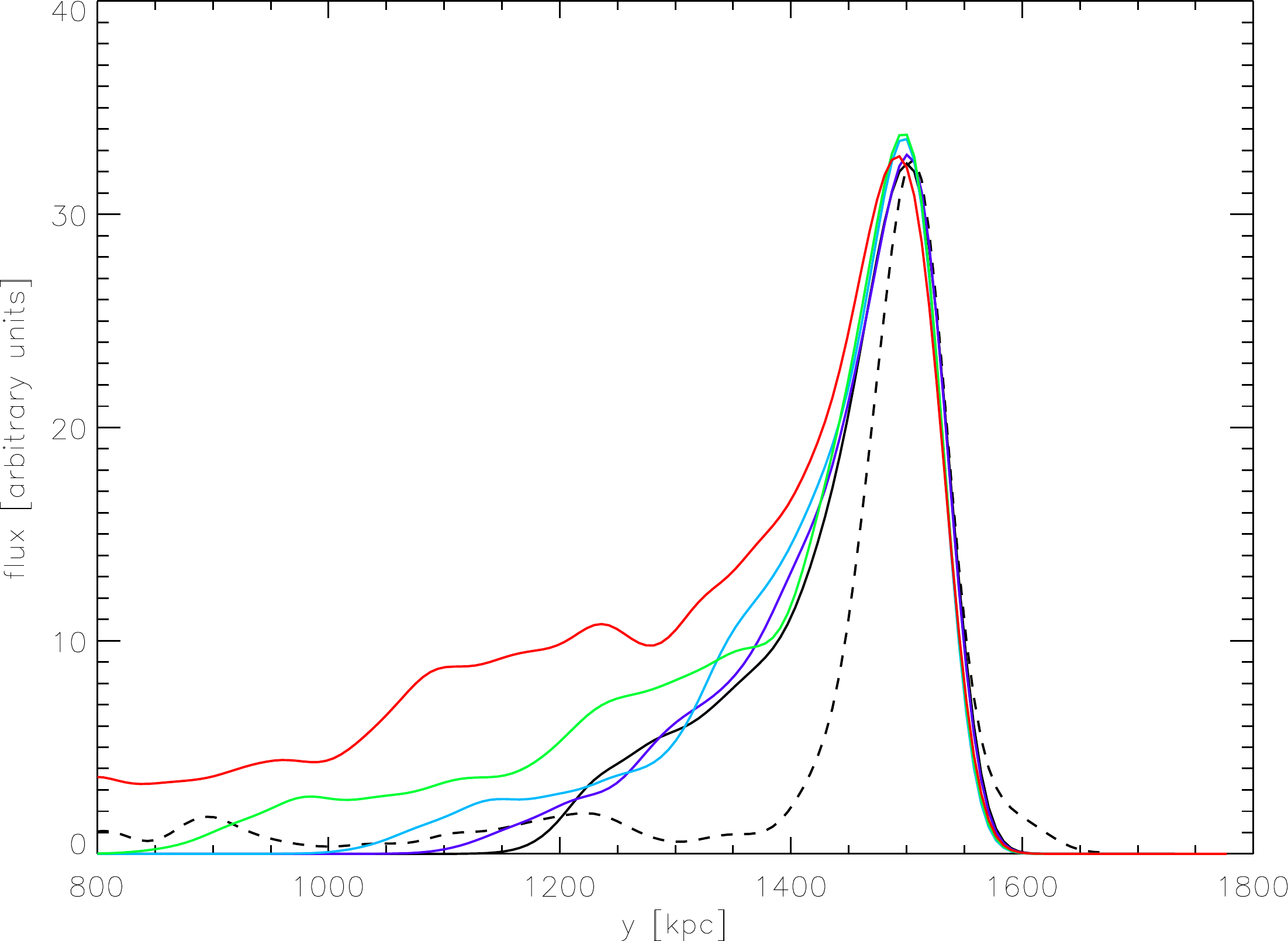}
\includegraphics[angle =90, trim =0cm 0cm 0cm 0cm,width=0.49\textwidth, clip=true]{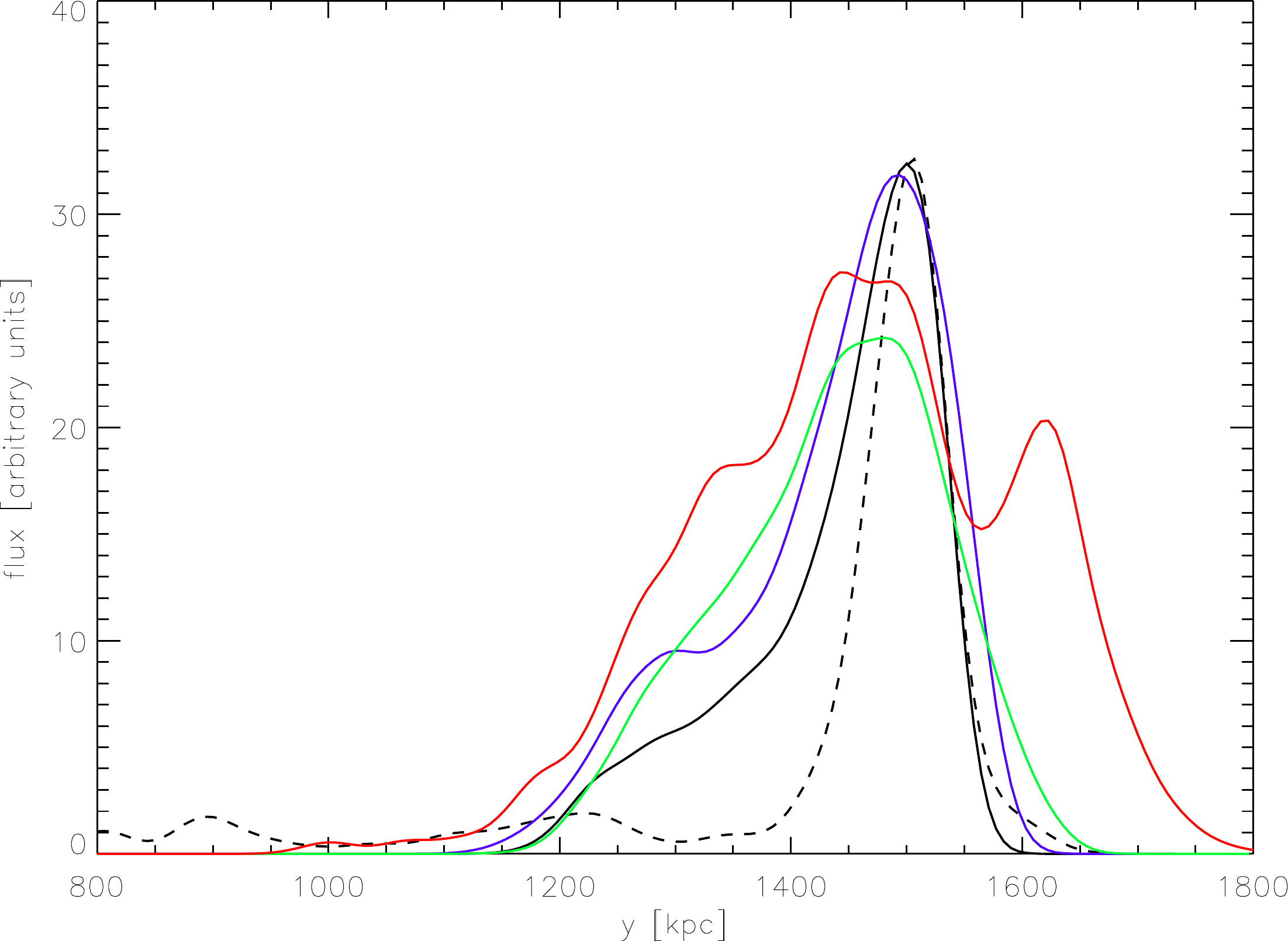}

\caption{Top: integrated brightness profiles across the y-axis for the top relic. Black dashed line displays the observed profile for the northern relic in CIZA J2242.8+5301.Solid lines, black: $b=0$~kpc, purple:  $b=1 r_{\rm{c}}$, blue: $b=2 r_{\rm{c}}$, green: $b=3 r_{\rm{c}}$, orange: $b=4 r_{\rm{c}}$, red: $b=5 r_{\rm{c}}$. Middle: integrated brightness profiles across the x-axis from $|x| < 100$~kpc for the top relic for relics seen under different viewing angles.  Black dashed line displays the observed profile for the northern relic in CIZA J2242.8+5301. Solid lines, black: $i=0\degr$, purple: $i=5\degr$, blue: $i=10\degr$, green: $i=20\degr$, red: $i=30\degr$. 
Bottom: integrated brightness profiles across the x-axis from $|x| < 100$~kpc for simulations including substructure. Black dashed line displays the observed profile for the northern relic in CIZA J2242.8+5301. Solid lines, black: no substructure, purple: $A=0.2$, $\lambda=100$~kpc, green: $A=0.3$, $\lambda=150$~kpc, red: $A=0.4$, $\lambda=200$~kpc. } 
\label{fig:profiles}
\end{figure}

Taking the $2:1$ merger event, we varied the impact parameter, see Fig.~\ref{fig:massratio}. As discussed in the beginning of Sect.~\ref{sec:results}, two asymmetric shock waves develop, breaking the cylinder symmetry (or reflection symmetry in the resulting image) around the merger axis.
 The variation in Mach number across the shock front and the amount of projected shock surface along the line of sight cause the radio emission to fade away more slowly on one side of the radio relics. Integrated brightness profiles for the top relic across the y-axis are shown in Fig.~\ref{fig:profiles} (top panel). The total extent of the relic decreases by about 10\% from $b=0$~kpc to $b=5 r_{\rm{c,1}}$.
 
 The CIZA J2242.8+5301 relics are more or less symmetrically located along the proposed merger axis, although the brightness profile at the east end of the northern relic is not well determined because of a bright compact radio source at this location. To obtain a better match with the observed radio map, we reflected the map along the y-axis, i.e., placing the off-axis merger point on the other side (this is equivalent of changing the impact parameter from $b$ to $-b$). The simulated profiles all peak around $x=0$~kpc, while the observed profile does not. From these profiles it is therefore difficult to constrain the impact parameter. However,  by looking at Fig.~\ref{fig:massratio} we find that $b \lesssim 3 r_{\rm{c,1}}$ is required to obtain a reasonable match.  For larger impact parameters the curvature of top shock does not match the observed one (i.e., this can be seen on the right side of the top relic in the R21b4 panel from  Fig.~\ref{fig:massratio}).

\subsubsection{Viewing angle}
\label{sec:viewangle}
We varied the viewing angle ($i$) by rotating our computational box around the x-axis before projecting it along the line of sight.  The extent of the emission towards the cluster center increases by about a factor of three from $i=5\degr$ to $i=30\degr$, see Fig.~\ref{fig:massratio}. Radial profiles of the brightness distribution are shown in Fig.~\ref{fig:profiles} (middle panel). The outer edge of the northern relic in CIZA~J2242.8+5301 is very pronounced and the brightness of the radio emission drops also quickly towards the cluster center, therefore the relic is likely seen close to edge on, roughly under an angle $\lesssim 10\deg$. In all cases, the simulated brightness profiles do not  reproduce the very quick drop towards the cluster center as it is observed for the northern relic.
The southern relic is more diffuse and extended in CIZA~J2242.8+5301, as is the case in our simulated maps.

\subsubsection{$\beta$-model}
\label{sec:beta}
\begin{figure*}
\includegraphics[angle =90, trim =0cm 0cm 0cm 0cm,width=1.0\textwidth, clip=true]{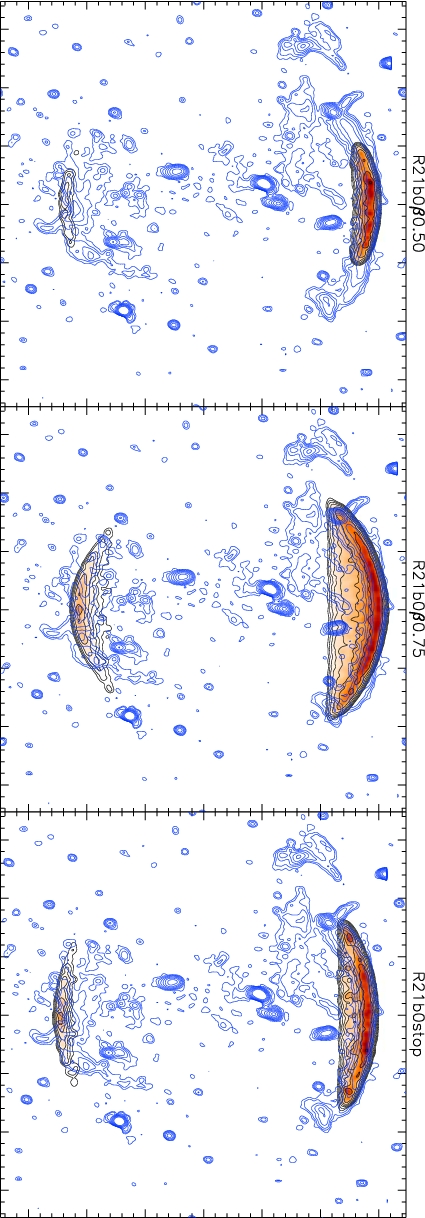}
\includegraphics[angle =90, trim =0cm 0cm 0cm 0cm,width=1.0\textwidth, clip=true]{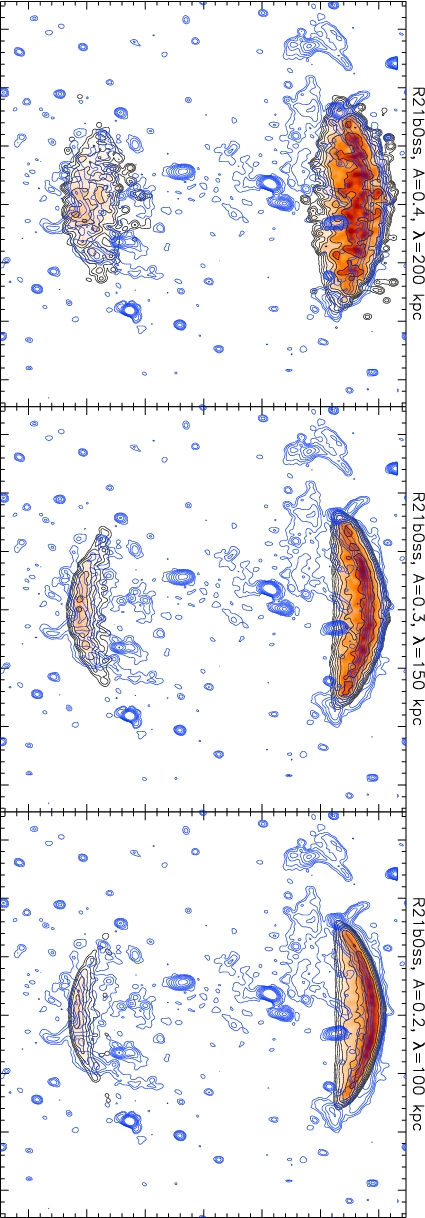}
\caption{Top row: Simulated radio emission for R21b0 cluster merger with different values of $\beta_1$ and $r_{\rm{c,1}}$ and  for a R21b0 cluster merger with the two halos sticking at core passage, which represents the extreme case of infinite tidal friction. 
Bottom row:  Simulated radio emission for a R21b0 cluster mergers with substructure. Contours and colors are the same as in Fig.~\ref{fig:massratio}. The images span $3.4 \times 3.4$~Mpc.  } 
\label{fig:beta}
\end{figure*}

One of the parameters which could influence the size of the radio relics is the core radius used in our $\beta$-profile (Eq.~\ref{eq:betamodel}). Because the total mass is fixed in our simulations the core radius is set by the $\beta$ parameter in Eq.~\ref{eq:betamodel}. We changed $\beta$ from the default value of $2/3$ to 0.5 and 0.75 for the most massive cluster to investigate the effect this has on the produced radio maps. For $\beta=0.5$, the core radius decreases from 134~kpc to about 83~kpc, while for $\beta=0.75$ the core radius increases to 163~kpc. We run the simulations with these different core radii and $\beta$ values for the standard R21b0 merger scenario. The resulting radio maps are displayed in Fig.~\ref{fig:beta}. 

The change in $r_{\rm{c,1}}$ has a direct influence on the size of the resulting shock waves and radio relics formed. For  $r_{\rm{c,1}}=83$~kpc, the northern and southern relics are almost a factor of 2 smaller in size. For $r_{\rm{c,1}}=163$~kpc,  the relics have a slightly larger size and are similar to the R1.51b0 merger (see Fig.~\ref{fig:massratio}). This implies that our best fitting mass ratio found in Sect.~\ref{sec:massratio} depend on the adopted core radii. In particular if the most massive cluster has a larger core radius this mimics a merger with a smaller mass ratio (e.g., $1.5:1$), see Fig.~\ref{fig:massratio}. 
By  increasing the core radius of the less massive subcluster the size of the smaller southern relic can be increased, to some extent mimicking a $2.5:1$ merger.
Taking different core radii, the range of matching mass ratios for  CIZA~J2242.8+5301 is therefore larger and lies roughly between $1.5:1$ and $2.5:1$ .

\subsubsection{Cool core}
\label{sec:coolcore}
In massive relaxed galaxy clusters the ICM temperature decreases strongly towards the center. To investigate the effect of such a cool core, we add a central cool core to the most massive subcluster using Eq.~\ref{eq:tcool}. We take typical value of $T_{\rm{min}}/T_{\rm{avg}}=0.3$, $r_{\rm{cool}} = 50$~kpc and  $r_{\rm{cool}}=2$ \citep[e.g.,][]{2006ApJ...640..691V} and use the standard R21b0 merger scenario. We find that the addition of a central cool core does not influence the size and shape of the resulting radio relics. The resulting radio map looks almost identical to the default R21b0 merger map, see Fig.~\ref{fig:massratio}.

\subsection{Spectral index}
\begin{figure}
\includegraphics[angle =90, trim =0cm 0cm 0cm 0cm,width=0.49\textwidth, clip=true]{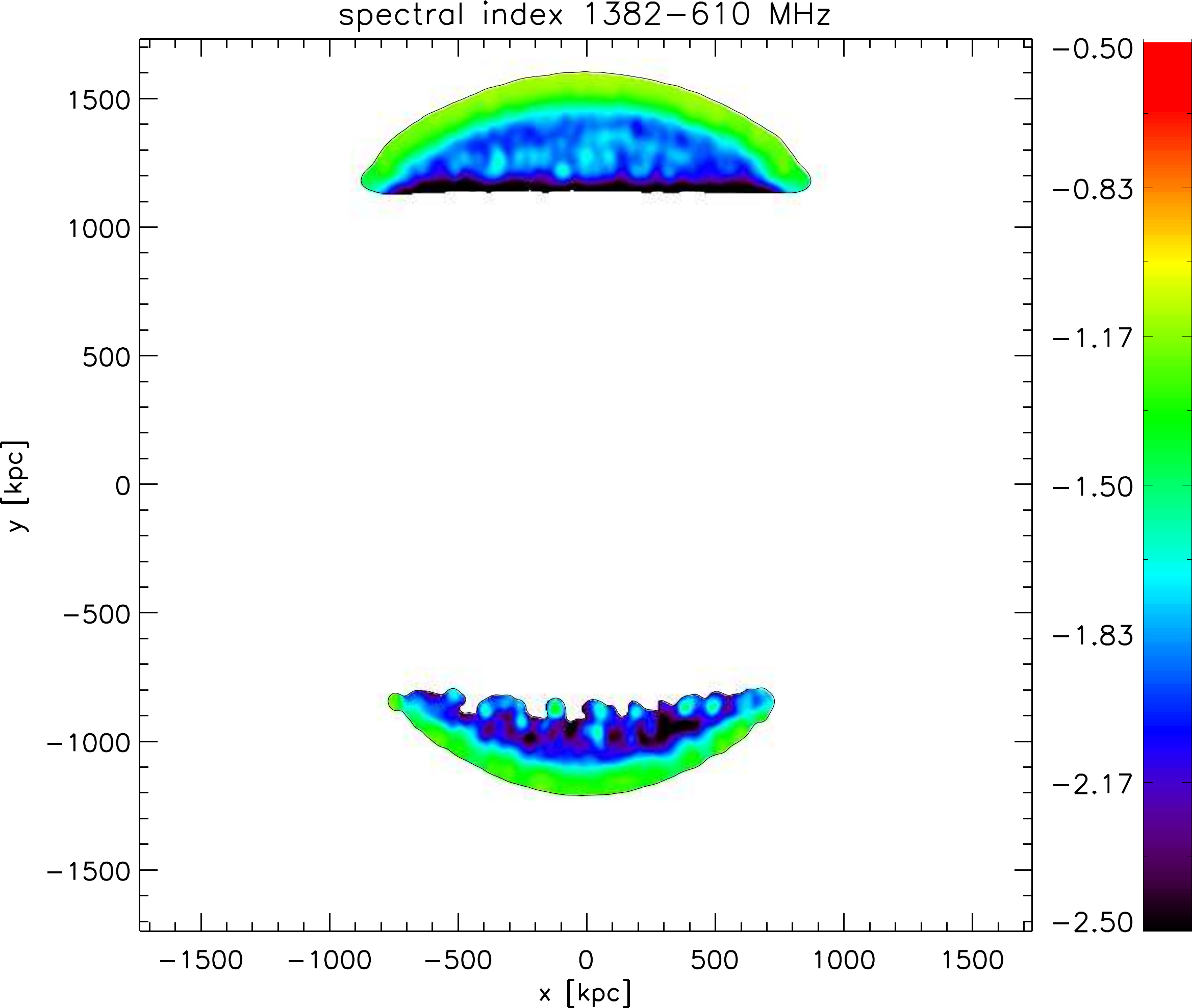}

\caption{Simulated spectral index map between 610 and 1382~MHz for the R21b0 merger. The small scale spectral index variations seen in the post-shock region are due to the finite number of particles used to compute the radio map.} 
\label{fig:spixr2b0}
\end{figure}

We computed a spectral index map between 610 and 1382~MHz for the R21b0 merger, again taking a constant magnetic strength of $5$~$\mu$Gauss, see Fig.~\ref{fig:spixr2b0}. We find that the spectral index in front of the two relics is $\sim-1.1$, while the spectral index steepens quickly to $\sim-2$ in the direction of the cluster center. After the initial spectral steepening to $\alpha\sim -2$, the spectral index remains constant. The radio emission in this region comes from (i) aged radio plasma with a steep spectrum in the post-shock region and (ii) projected radio emission from the front of the shock wave with a flatter spectral index of $\sim -1$.

The observed spectral index in front of the northern CIZA~J2242.8+5301 relic is flatter, with $\alpha \sim -0.6 $ or $-0.7$, than in our simulated spectral index map. The observed spectral index corresponds to a $\mathcal{M} = 4.6$ shock. In the our R21b0 merger, we find a Mach number of about 2.7--2.4 for the upper shock wave. This translates into $\alpha_{\rm{inj}}=-0.8$ to $-0.9$. However, in the simulated map $\alpha$ is steeper by about 0.3 units at the front of the relic. This difference is caused by the limited spatial resolution ($\sim50$~kpc) of our synthetic radio map: radio emission directly at the front of the shock with $\alpha_{\rm{inj}}\approx-0.85$ is always mixed with some emission from particles that have already have undergone some synchrotron losses. For CIZA~J2242.8+5301 this implies that the true injection spectral index is probably even flatter than $-0.6$, if the shock front has a similar shape as in our simulations.

\section{Substructure \& clumping}
\label{sec:substructure}
Real mergers are more complex than in our idealized simulations. Hence, the detailed properties, such as the shapes or brightness profiles of radio relics, do not necessarily match with the observations. In particular the presence of less massive substructures and galaxy filaments extending from the large scale cosmic web will modify the shock morphologies and locations \citep[e.g.,][]{2010arXiv1001.1170P}. Indeed, there are additional smaller relics located in CIZA~J2242.8+5301 that points towards a more complex merger event. These more complex relics are also seen in simulations which include large-scale structure formation with radio emission from shocks \citep[e.g.,][]{2008MNRAS.385.1211P, 2008MNRAS.391.1511H,2009MNRAS.393.1073B}. Although we cannot easily embed our simulation into a large-scale cosmological environment, we investigate the effect of small scale ($\lesssim 200$~kpc) substructures/clumps. 
Matter falling onto clusters is expected to be clumpy in nature and the clumps have typical dimensions of galaxies or small galaxy-groups \citep{2006MNRAS.373.1339R}.
If the gas in the cluster's outskirts is clumpy, and this is not taken into account,  it can lead to an overestimation of the baryon fraction and affect the derived entropy and pressure profiles \citep[e.g., ][]{2011Sci...331.1576S}.  The presence of well-defined radio relics with sharp outer rims can be used to constrain the amount of clumping in the ICM at large distances ($\gtrsim 1$~Mpc) from the cluster center.

\begin{figure}
\includegraphics[angle =90, trim =0cm 0cm 0cm 0cm,width=0.49\textwidth, clip=true]{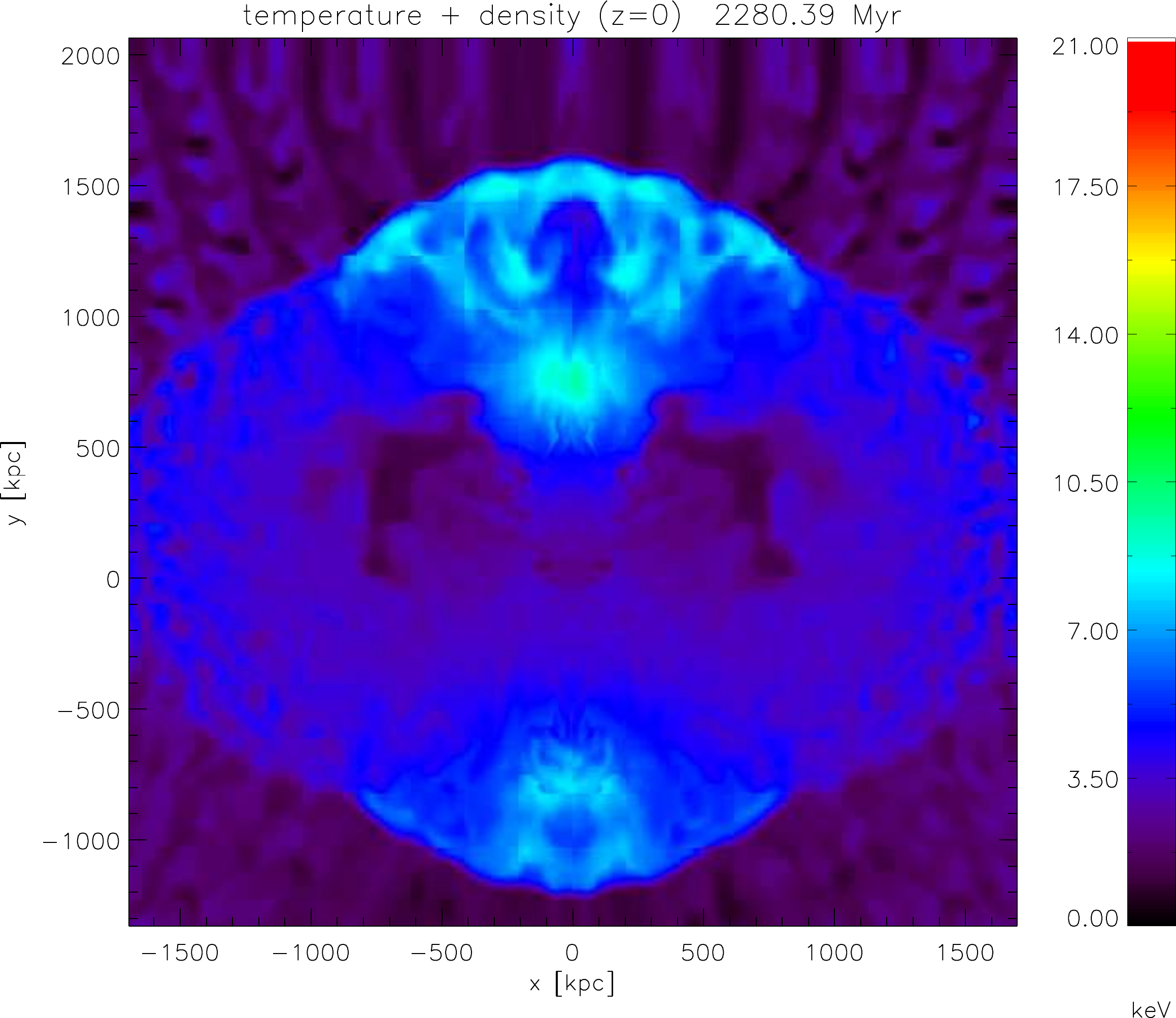}
\caption{Left: Temperature distribution at the slice $z=0$ for the R21b0ss (A=0.4, $\lambda=200$~kpc) merger.} 
\label{fig:lumpy40_200}
\end{figure}

We modified our initial density and temperature profiles by adding sinusoidal density fluctuations throughout the computational volume. The two parameters which control this are the relative amplitude ($A$) and wavelength ($\lambda$) of these density fluctuations. These sinusoidal waves were inserted parallel to the x, y and z-axes.
To keep the pressure unchanged, we also included temperature fluctuations that are inversely proportional to the density fluctuations, see Fig.~\ref{fig:lumpy40_200} for an example.  We carried out several runs changing the amplitude and wavelength of the fluctuations. Radio images from runs with $A=0.2, 0.3, 0.4$ and $\lambda=100, 150, 200$~kpc, respectively, are shown in Fig.~\ref{fig:beta}. Brightness profiles along the y-axis are displayed in Fig.~\ref{fig:profiles} (bottom panel). 

The resulting radio relics are similar in shape to the canonical R21b0 merger map, but the relics are wider and display brightness fluctuations in the radio emission. The effect of this substructure is that the outer boundary of the relics is less well defined. For relics observed close to edge-on, projection effects average out some of the variations in the strength of the radio emission. 
The northern relic in CIZA~J2242.8+5301 has a smooth appearance and well-defined outer rim. This probably implies that few clumps are present this location in the ICM. We find that the radio maps with $A\gtrsim 0.3$ do not provide a good match to the observed northern relic in CIZA~J2242.8+5301. There is little difference 
between runs with $A= 0.3$ and $\lambda=75, 150, 200$~kpc. 
We can therefore put upper limits on the amplitude of density fluctuations of roughly 30\%, for substructures with scales of $\lesssim 200$~kpc, at the location of the  northern relic. This means that clumping appears to be much less than that inferred from Suzaku observations of Perseus with $\sqrt{<n_e^2>/<n_e>^2}\sim 3-4$ near the virial radius \citep[e.g., ][]{2011Sci...331.1576S}. The reason for this could be that the shock waves in CIZA~J2242.8+5301 are expanding into a region sufficiently far from filaments such that it is largely free from infalling substructure.

A more realistic representation of density clumps could probably be used to put tighter constraints on the amount of substructure. This is beyond the scope of this paper and we leave this for future work.

\section{Discussion}
\label{sec:discussion}
\subsection{Effect of dark matter dynamics}
\label{sec:dm}
Clearly, a gross simplification of our setup is the treatment of the gravity in the form of rigid potentials that move on set trajectories without interacting. In reality, the dark matter particles of the progenitor halos will interact in the course of the merger to form a new halo. In the course of this, the potential changes mainly due to two effects: i) dynamic friction and ii) tidal shocking. The first one is the most important effect. After core passage, it will cause the halos to slow with respect to each other until they turn around on their orbits. As the merger shock waves are running out of the cluster faster than the speeds of the halos with respect to each other, this will not change the results dramatically. To test the magnitude of this effect, we ran a simulation where we made the two halos stick at core passage, which represents the extreme case of infinite tidal friction. The result is shown in Fig.~\ref{fig:beta}.  The relics are now somewhat narrower (about 10\% for the top relic) but apart from that  there are not many changes in the radio map even though the temperature and density are quite different in the center. The main
change are the lower Mach numbers (which drop from 2.7 to 2.2, and from 2.2 to 1.9 for the top and bottom relics compared to the default R21b0 merger), but since this affects both for the top and bottom shocks the end result is not too different from the default R21b0 map. Finally, we ignore the effect of tidal shocking, which leads to heating of the central DM orbits of the cluster and leads to a puffing up of the potential. This will affect the cluster potential within the core radius but will leave the regions where the shocks are located largely unaffected. For a proper simulation of this effect, one would need to resort to a simulation with a dynamic DM component.

\subsection{Relic width and brightness profiles}
One of the most intriguing properties of the northern relic in CIZA~J2242.8+5301 is its very narrow width and rather uniform luminosity along its extent. From our simulated radio map we  find the brightest part of the top relic to be $\sim20\%$ wider than the observed width. In Sect. \ref{sec:dm} we found a smaller width by including infinite tidal friction. This affects the shock downstream velocity, which determines how far the the synchrotron emitting  electrons are carried away from the shock front \citep[e.g., ][]{2005ApJ...627..733M}. The relic's width is linearly proportional to the downstream shock velocity. In case of  infinite tidal friction the simulated radio relic is only $\sim10\%$ wider compared to the observed width. In addition, the width depends on the adopted magnetic field strength, but increasing $B$ to 7~$\mu$Gauss decreases the width only by a few percent. This implies that projection effects and the finite resolution of our simulations play and important role in determining the width.

For the southern relic, the emission fades  slowly towards the east and west ends. The reason why this does not happen for the top relic in the simulation is because of the quick drop in the Mach number at $|x| > 850$~kpc, see Fig.~\ref{fig:vfield} (left panel). However, for the CIZA~J2242.8+5301 northern relic the integrated brightness profile across the y-axis (Fig~\ref{fig:profiles}, top panel) drops much quicker and is flatter than in the simulations. The reason for this is unclear, it could be caused by variations in the magnetic field strength, Mach number, or because the shock surface is more complex. The Mach number found from the injection spectral index is higher than that in the simulations (4.6 versus 2.7 for the northern relic), this difference is puzzling. It could indicate that our temperature profiles are not correct or that Eq.~\ref{eq:inj-mach}, which links $\alpha_{\rm{inj}}$ and $\mathcal{M}$, is not directly applicable. X-ray observations are crucial here to directly determine the Mach number from the temperature jump.

\subsection{Origin of single radio relics}

Another interesting outcome of our simulations concerns the existence of single radio relics, e.g., the relic in A521 \citep{2008A&A...486..347G}. In binary cluster mergers, two shocks should be generated and hence one expects the majority of relics to come in pairs. In the simulations with mass ratios $\gtrsim 3$ very little radio emission is generated by the shock wave in front of the less massive subcluster. The Mach number of these shocks is quite low ($\mathcal{M} \lesssim 2$) and the shock surfaces are small, which results in very little observable radio emission.  Based on the simulations, one would thus expect most double relics to be located in merging systems with small mass ratios, while single should be found in mergers with larger mass ratios.

\subsection{A quantitative metric for the goodness of fit}

We did not define a quantitative metric for the goodness of fit between the simulated and observed radio images. This paper is meant to be illustrative of the potential of the method and in this work we chose to apply it only to the relics in CIZA~J2242.8+5301. We also note that the position, shape, and
surface brightness distribution of relics are influenced by the cluster environment (e.g., connected galaxy filaments) which affect the derived merger parameters. However, for future work when constraining merger parameters for larger numbers of double relics systems, it is important to develop a metric for the goodness of fit.  We discuss here a possible approach and address some of the issues involved.

 A first step should be the removal (or masking) of point sources and extended radio galaxies. In addition, one may have to subtract the emission from a  radio halo in the cluster center. Removing extended radio galaxies and a radio halo is not entirely trivial as the spatial scales of these structures can be comparable to that of relics. However, this should still be feasible as double radio relics are relatively bright and their morphology is quite different from radio halos or radio galaxies. 
 
For constraining the merger parameters there are two aspects that can be considered: (i) the spatial information of the location of the relics and (ii) the brightness distribution. We ignore information about the spectral and polarization properties because this complicates matters and this information is not available for all double relics. Due to the uncertainties in the magnetic field strength the brightness of the emission is probably not a very good criteria for a quantitative metric. It is also not entirely clear under what conditions relics precisely form, i.e, in the case of Abell 2146 two shocks are found in a Chandra X-ray image, but no bright radio relics are detected \citep{2010MNRAS.406.1721R, 2011arXiv1105.0435R}.

Radio relics, especially their outer edges, should directly trace the location of the merger shock waves and thus provide most of the constraints for the merger scenario. This is also what we mainly used in our qualitative  comparison for the relics in CIZA~J2242.8+5301. The viewing geometry is mostly constrained from the downstream extent of the relics. In the light of what we discussed above, we propose a two-dimensional spatial cross-correlation.  This should be repeated for every possible rotation angle for a given image from the simulations. The height of the peak in the cross-correlation then gives a quantitative metric for the goodness of fit for a single simulated image. Image scaling should not be considered here as this relates to the absolute sizes and distances of the relics which evolve over time, and the time since core passage is one of the parameters we wish to constrain. 
The process can then be repeated for different merger parameters (mass ratios and impact parameters),  snapshots in time and viewing angles to find the best fit.  Because the brightness distribution is uncertain, we propose to apply a fixed cut in the radio flux (a few times the noise level in the observed radio image), and put all pixels values to either $1$ (emission) or $0$ (no emission). This is done also for the simulated images. The overall normalization of the simulated radio emission can be fixed by normalizing to the total flux in the observed image. Instead of a cross-correlation, one could also consider a simple image subtraction (after x, y-shifting and rotation) followed by taking the sum of the absolute difference. It may be worth giving the outer edges of the relics a somewhat large weight in the process, for example by setting these to a higher value (e.g., $2$), some experimentation will be needed here. 
Applications of these cross-correlations shall be reserved for future work.

\section{Summary}
\label{sec:summary}
We have presented simulated radio maps of idealized binary cluster mergers, varying the mass ratios, impact parameters, viewing angles and density/temperature profiles. The radio emission in these simulations is produced by relativistic electrons which are accelerated via the DSA mechanism in the shock waves created by the merger event. 

We have used these simulated radio maps to constrain the merger scenario for the cluster CIZA~J2242.8+5301, which hosts a Mpc scale double radio relic. We find that a merger with a mass ratio between $1.5:1$ and $2.5:1$ provides the best match to the observed radio emission in this cluster. The main uncertainty in the derived mass ratio comes from the unknown core radii of the clusters before the merger. The viewing angle is mostly constrained by the extent of the top relic towards the cluster center,  additional substructure/clumps also increase the relic's extent in the downstream region. Since the observed extent is only 55~kpc this implies that the relic is seen close to edge-on (i.e., $i\lesssim10\degr$). The impact parameter is difficult to constrain as the observed brightness profile along the x-axis is not reproduced for any our merger scenarios. Only for $b\gtrsim 3r_{\rm{c,1}} \approx 400$~kpc we find that the curvature of the simulated relics do not match with the observations.
From our simulations we find that the time since core passage does not depend much on the adopted initial conditions. We therefore constrain this number to $1.0$~Gyr,  although it also depends on the adopted total mass of the system and temperature profile.

One of the main uncertainties for comparing the simulated radio emission with observations, in particular the brightness profiles, is the structure and strength of the magnetic field. Especially the brightness profile of the main relic along the relic's length is not well reproduced by our simulations. Little is known about the magnetic field strength at distances larger than 1~Mpc from the cluster center. Using equipartition arguments magnetic fields strengths of the order of $\sim 1$~$\mu$G have been derived  \citep[e.g.,][]{2004IJMPD..13.1549G}. Upper limits on the IC emission from radio relics indicate magnetic fields of a few $\mu$G \citep[e.g.,][]{2009PASJ...61..339N, 2010arXiv1004.2331F}. Faraday rotation measures give similar results \citep[e.g.,][]{2010A&A...513A..30B}. However, the spatial variations of the magnetic field strength at these distances from the cluster center are unknown.  In addition, several smaller relics are found in the cluster which are not present in our simulations. These relics  are probably caused by additional substructures which are not included in our initial conditions.

We find that radio relics can be used to put constraints of the amount of clumping or substructure in the ICM at large distances ($\sim$ 1 -- 2~Mpc) from the cluster center, under the assumption that relics are tracers of shocks. The sharp outer boundary of the northern relic in CIZA~J2242.8+5301 implies density fluctuations with an amplitude $\lesssim 30\%$ of the average density (the size of fluctuations being $\sim 200$~kpc).

The simulations also indicate that most double relics should be located in merging system with mass ratios between $1:1$ and $1:3$, while single radio relics from in clusters that undergo less energetic mergers with larger mass ratios. 

We conclude that double radio relics can be used to constrain cluster merger events, which is especially useful when detailed X-ray observations are not available. Relics are often located at large distances from the cluster center, where the density is so low that even deep X-ray observations cannot be used to probe the physical conditions there (e.g., to measure the temperature jump). In addition, relics can be located in lower-mass systems where the thermal X-ray emission is not detected in the ROSAT All Sky Survey, for example \citep{1999A&A...349..389V}. Some examples of possible radio relics in poor systems without a X-ray counterparts have been reported by \cite{2009AJ....137.3158B, 2009A&A...508...75V, 2011A&A...527A.114V}.

One important parameter is the strength and structure of the magnetic field at the location of the radio relics. MHD simulations of cluster mergers will be needed to study the variations and strength  of the magnetic field at large distances from the cluster center, and investigate the effect this has on the derived merger parameters from radio maps.

\section*{Acknowledgments}
We would like to thank the referee Shea Brown for his comments and useful suggestions. 
RJvW acknowledges funding from the Royal Netherlands Academy of Arts and Sciences. MB and MH acknowledge support by the research group FOR 1254 funded by the Deutsche Forschungsgemeinschaft. We gratefully acknowledge the use of the supercomputers JUROPA at the John von Neumann Institute for Computing in Juelich. The software used in this work was in part developed by the DOE-supported ASC / Alliance Center for Astrophysical Thermonuclear Flashes at the University of Chicago.

\bibliography{../ref_filaments_mnras.bib}

\begin{thebibliography}{}

\bibitem[\protect\citeauthoryear{{Akahori} \& {Yoshikawa}}{{Akahori} \&
  {Yoshikawa}}{2010}]{2010PASJ...62..335A}
{Akahori} T.,  {Yoshikawa} K.,  2010, \pasj, 62, 335

\bibitem[\protect\citeauthoryear{{Allen}, {Schmidt} \& {Fabian}}{{Allen}
  et~al.}{2001}]{2001MNRAS.328L..37A}
{Allen} S.~W.,  {Schmidt} R.~W.,    {Fabian} A.~C.,  2001, \mnras, 328, L37

\bibitem[\protect\citeauthoryear{{Axford}, {Leer} \& {Skadron}}{{Axford}
  et~al.}{1977}]{1977ICRC...11..132A}
{Axford} W.~I.,  {Leer} E.,    {Skadron} G.,  1977, in International Cosmic Ray
  Conference Vol.~11 of International Cosmic Ray Conference, {The Acceleration
  of Cosmic Rays by Shock Waves}.
pp 132--+

\bibitem[\protect\citeauthoryear{{Bagchi}, {Durret}, {Neto} \& {Paul}}{{Bagchi}
  et~al.}{2006}]{2006Sci...314..791B}
{Bagchi} J.,  {Durret} F.,  {Neto} G.~B.~L.,    {Paul} S.,  2006, Science, 314,
  791

\bibitem[\protect\citeauthoryear{{Bagchi}, {Sirothia}, {Werner}, {Pandge},
  {Kantharia}, {Ishwara-Chandra}, {Gopal-Krishna}, {Paul} \& {Joshi}}{{Bagchi}
  et~al.}{2011}]{2011ApJ...736L...8B}
{Bagchi} J.,  {Sirothia} S.~K.,  {Werner} N.,  {Pandge} M.~B.,  {Kantharia}
  N.~G.,  {Ishwara-Chandra} C.~H.,  {Gopal-Krishna} {Paul} S.,    {Joshi} S.,
  2011, \apjl, 736, L8+

\bibitem[\protect\citeauthoryear{{Battaglia}, {Pfrommer}, {Sievers}, {Bond} \&
  {En{\ss}lin}}{{Battaglia} et~al.}{2009}]{2009MNRAS.393.1073B}
{Battaglia} N.,  {Pfrommer} C.,  {Sievers} J.~L.,  {Bond} J.~R.,
  {En{\ss}lin} T.~A.,  2009, \mnras, 393, 1073

\bibitem[\protect\citeauthoryear{{Bautz}, {Miller}, {Sanders}, {Arnaud},
  {Mushotzky}, {Porter}, {Hayashida}, {Henry}, {Hughes}, {Kawaharada},
  {Makashima}, {Sato} \& {Tamura}}{{Bautz} et~al.}{2009}]{2009PASJ...61.1117B}
{Bautz} M.~W.,  {Miller} E.~D.,  {Sanders} J.~S.,  {Arnaud} K.~A.,  {Mushotzky}
  R.~F.,  {Porter} F.~S.,  {Hayashida} K.,  {Henry} J.~P.,  {Hughes} J.~P.,
  {Kawaharada} M.,  {Makashima} K.,  {Sato} M.,    {Tamura} T.,  2009, \pasj,
  61, 1117

\bibitem[\protect\citeauthoryear{{Bell}}{{Bell}}{1978a}]{1978MNRAS.182..147B}
{Bell} A.~R.,  1978a, \mnras, 182, 147

\bibitem[\protect\citeauthoryear{{Bell}}{{Bell}}{1978b}]{1978MNRAS.182..443B}
{Bell} A.~R.,  1978b, \mnras, 182, 443

\bibitem[\protect\citeauthoryear{{Blandford} \& {Eichler}}{{Blandford} \&
  {Eichler}}{1987}]{1987PhR...154....1B}
{Blandford} R.,  {Eichler} D.,  1987, \physrep, 154, 1

\bibitem[\protect\citeauthoryear{{Blandford} \& {Ostriker}}{{Blandford} \&
  {Ostriker}}{1978}]{1978ApJ...221L..29B}
{Blandford} R.~D.,  {Ostriker} J.~P.,  1978, \apjl, 221, L29

\bibitem[\protect\citeauthoryear{{B{\"o}hringer}, {Voges}, {Huchra}, {McLean},
  {Giacconi}, {Rosati}, {Burg}, {Mader}, {Schuecker}, {Simi{\c c}}, {Komossa},
  {Reiprich}, {Retzlaff} \& {Tr{\"u}mper}}{{B{\"o}hringer}
  et~al.}{2000}]{2000ApJS..129..435B}
{B{\"o}hringer} H.,  {Voges} W.,  {Huchra} J.~P.,  {McLean} B.,  {Giacconi} R.,
   {Rosati} P.,  {Burg} R.,  {Mader} J.,  {Schuecker} P.,  {Simi{\c c}} D.,
  {Komossa} S.,  {Reiprich} T.~H.,  {Retzlaff} J.,    {Tr{\"u}mper} J.,  2000,
  \apjs, 129, 435

\bibitem[\protect\citeauthoryear{{Bonafede}, {Feretti}, {Murgia}, {Govoni},
  {Giovannini}, {Dallacasa}, {Dolag} \& {Taylor}}{{Bonafede}
  et~al.}{2010}]{2010A&A...513A..30B}
{Bonafede} A.,  {Feretti} L.,  {Murgia} M.,  {Govoni} F.,  {Giovannini} G.,
  {Dallacasa} D.,  {Dolag} K.,    {Taylor} G.~B.,  2010, \aap, 513, A30+

\bibitem[\protect\citeauthoryear{{Bonafede}, {Giovannini}, {Feretti}, {Govoni}
  \& {Murgia}}{{Bonafede} et~al.}{2009}]{2009A&A...494..429B}
{Bonafede} A.,  {Giovannini} G.,  {Feretti} L.,  {Govoni} F.,    {Murgia} M.,
  2009, \aap, 494, 429

\bibitem[\protect\citeauthoryear{{Brown}, {Duesterhoeft} \& {Rudnick}}{{Brown}
  et~al.}{2011}]{2011ApJ...727L..25B}
{Brown} S.,  {Duesterhoeft} J.,    {Rudnick} L.,  2011, \apjl, 727, L25+

\bibitem[\protect\citeauthoryear{{Brown} \& {Rudnick}}{{Brown} \&
  {Rudnick}}{2009}]{2009AJ....137.3158B}
{Brown} S.,  {Rudnick} L.,  2009, \aj, 137, 3158

\bibitem[\protect\citeauthoryear{{Burns}, {Roettiger}, {Ledlow} \&
  {Klypin}}{{Burns} et~al.}{1994}]{1994ApJ...427L..87B}
{Burns} J.~O.,  {Roettiger} K.,  {Ledlow} M.,    {Klypin} A.,  1994, \apjl,
  427, L87

\bibitem[\protect\citeauthoryear{{Cavaliere} \& {Fusco-Femiano}}{{Cavaliere} \&
  {Fusco-Femiano}}{1976}]{1976A&A....49..137C}
{Cavaliere} A.,  {Fusco-Femiano} R.,  1976, \aap, 49, 137

\bibitem[\protect\citeauthoryear{{Clowe}, {Brada{\v c}}, {Gonzalez},
  {Markevitch}, {Randall}, {Jones} \& {Zaritsky}}{{Clowe}
  et~al.}{2006}]{2006ApJ...648L.109C}
{Clowe} D.,  {Brada{\v c}} M.,  {Gonzalez} A.~H.,  {Markevitch} M.,  {Randall}
  S.~W.,  {Jones} C.,    {Zaritsky} D.,  2006, \apjl, 648, L109

\bibitem[\protect\citeauthoryear{{Croston}, {Pratt}, {B{\"o}hringer}, {Arnaud},
  {Pointecouteau}, {Ponman}, {Sanderson}, {Temple}, {Bower} \&
  {Donahue}}{{Croston} et~al.}{2008}]{2008A&A...487..431C}
{Croston} J.~H.,  {Pratt} G.~W.,  {B{\"o}hringer} H.,  {Arnaud} M.,
  {Pointecouteau} E.,  {Ponman} T.~J.,  {Sanderson} A.~J.~R.,  {Temple} R.~F.,
  {Bower} R.~G.,    {Donahue} M.,  2008, \aap, 487, 431

\bibitem[\protect\citeauthoryear{{Drury}}{{Drury}}{1983}]{1983RPPh...46..973D}
{Drury} L.~O.,  1983, Reports on Progress in Physics, 46, 973

\bibitem[\protect\citeauthoryear{{Ensslin}, {Biermann}, {Klein} \&
  {Kohle}}{{Ensslin} et~al.}{1998}]{1998A&A...332..395E}
{Ensslin} T.~A.,  {Biermann} P.~L.,  {Klein} U.,    {Kohle} S.,  1998, \aap,
  332, 395

\bibitem[\protect\citeauthoryear{{Fabian}, {Nulsen} \& {Canizares}}{{Fabian}
  et~al.}{1991}]{1991A&ARv...2..191F}
{Fabian} A.~C.,  {Nulsen} P.~E.~J.,    {Canizares} C.~R.,  1991, \aapr, 2, 191

\bibitem[\protect\citeauthoryear{{Finoguenov}, {Sarazin}, {Nakazawa}, {Wik} \&
  {Clarke}}{{Finoguenov} et~al.}{2010}]{2010arXiv1004.2331F}
{Finoguenov} A.,  {Sarazin} C.~L.,  {Nakazawa} K.,  {Wik} D.~R.,    {Clarke}
  T.~E.,  2010, ArXiv e-prints

\bibitem[\protect\citeauthoryear{{Fryxell}, {Olson}, {Ricker}, {Timmes},
  {Zingale}, {Lamb}, {MacNeice}, {Rosner}, {Truran} \& {Tufo}}{{Fryxell}
  et~al.}{2000}]{2000ApJS..131..273F}
{Fryxell} B.,  {Olson} K.,  {Ricker} P.,  {Timmes} F.~X.,  {Zingale} M.,
  {Lamb} D.~Q.,  {MacNeice} P.,  {Rosner} R.,  {Truran} J.~W.,    {Tufo} H.,
  2000, \apjs, 131, 273

\bibitem[\protect\citeauthoryear{{George}, {Fabian}, {Sanders}, {Young} \&
  {Russell}}{{George} et~al.}{2009}]{2009MNRAS.395..657G}
{George} M.~R.,  {Fabian} A.~C.,  {Sanders} J.~S.,  {Young} A.~J.,    {Russell}
  H.~R.,  2009, \mnras, 395, 657

\bibitem[\protect\citeauthoryear{{Giacintucci}, {Venturi}, {Macario},
  {Dallacasa}, {Brunetti}, {Markevitch}, {Cassano}, {Bardelli} \&
  {Athreya}}{{Giacintucci} et~al.}{2008}]{2008A&A...486..347G}
{Giacintucci} S.,  {Venturi} T.,  {Macario} G.,  {Dallacasa} D.,  {Brunetti}
  G.,  {Markevitch} M.,  {Cassano} R.,  {Bardelli} S.,    {Athreya} R.,  2008,
  \aap, 486, 347

\bibitem[\protect\citeauthoryear{{Govoni} \& {Feretti}}{{Govoni} \&
  {Feretti}}{2004}]{2004IJMPD..13.1549G}
{Govoni} F.,  {Feretti} L.,  2004, International Journal of Modern Physics D,
  13, 1549

\bibitem[\protect\citeauthoryear{{Hernquist}}{{Hernquist}}{1987}]{1987ApJS...6%
4..715H}
{Hernquist} L.,  1987, \apjs, 64, 715

\bibitem[\protect\citeauthoryear{{Hoeft} \& {Br{\"u}ggen}}{{Hoeft} \&
  {Br{\"u}ggen}}{2007}]{2007MNRAS.375...77H}
{Hoeft} M.,  {Br{\"u}ggen} M.,  2007, \mnras, 375, 77

\bibitem[\protect\citeauthoryear{{Hoeft}, {Br{\"u}ggen}, {Yepes},
  {Gottl{\"o}ber} \& {Schwope}}{{Hoeft} et~al.}{2008}]{2008MNRAS.391.1511H}
{Hoeft} M.,  {Br{\"u}ggen} M.,  {Yepes} G.,  {Gottl{\"o}ber} S.,    {Schwope}
  A.,  2008, \mnras, 391, 1511

\bibitem[\protect\citeauthoryear{{Hoshino}, {Henry}, {Sato}, {Akamatsu},
  {Yokota}, {Sasaki}, {Ishisaki}, {Ohashi}, {Bautz}, {Fukazawa}, {Kawano},
  {Furuzawa}, {Hayashida}, {Tawa}, {Hughes}, {Kokubun} \& {Tamura}}{{Hoshino}
  et~al.}{2010}]{2010PASJ...62..371H}
{Hoshino} A.,  {Henry} J.~P.,  {Sato} K.,  {Akamatsu} H.,  {Yokota} W.,
  {Sasaki} S.,  {Ishisaki} Y.,  {Ohashi} T.,  {Bautz} M.,  {Fukazawa} Y.,
  {Kawano} N.,  {Furuzawa} A.,  {Hayashida} K.,  {Tawa} N.,  {Hughes} J.~P.,
  {Kokubun} M.,    {Tamura} T.,  2010, \pasj, 62, 371

\bibitem[\protect\citeauthoryear{{Jaffe} \& {Perola}}{{Jaffe} \&
  {Perola}}{1973}]{1973A&A....26..423J}
{Jaffe} W.~J.,  {Perola} G.~C.,  1973, \aap, 26, 423

\bibitem[\protect\citeauthoryear{{Jones} \& {Ellison}}{{Jones} \&
  {Ellison}}{1991}]{1991SSRv...58..259J}
{Jones} F.~C.,  {Ellison} D.~C.,  1991, Space Science Reviews, 58, 259

\bibitem[\protect\citeauthoryear{{Kardashev}}{{Kardashev}}{1962}]{1962SvA.....%
6..317K}
{Kardashev} N.~S.,  1962, Soviet Astronomy, 6, 317

\bibitem[\protect\citeauthoryear{{Kempner}, {Blanton}, {Clarke}, {En{\ss}lin},
  {Johnston-Hollitt} \& {Rudnick}}{{Kempner}
  et~al.}{2004}]{2004rcfg.proc..335K}
{Kempner} J.~C.,  {Blanton} E.~L.,  {Clarke} T.~E.,  {En{\ss}lin} T.~A.,
  {Johnston-Hollitt} M.,    {Rudnick} L.,  2004, in {Reiprich} T.,  {Kempner}
  J.,   {Soker} N.,  eds, The Riddle of Cooling Flows in Galaxies and Clusters
  of galaxies {Conference Note: A Taxonomy of Extended Radio Sources in
  Clusters of Galaxies}.
pp 335--+

\bibitem[\protect\citeauthoryear{{Keshet}, {Waxman} \& {Loeb}}{{Keshet}
  et~al.}{2004}]{2004ApJ...617..281K}
{Keshet} U.,  {Waxman} E.,    {Loeb} A.,  2004, \apj, 617, 281

\bibitem[\protect\citeauthoryear{{Keshet}, {Waxman}, {Loeb}, {Springel} \&
  {Hernquist}}{{Keshet} et~al.}{2003}]{2003ApJ...585..128K}
{Keshet} U.,  {Waxman} E.,  {Loeb} A.,  {Springel} V.,    {Hernquist} L.,
  2003, \apj, 585, 128

\bibitem[\protect\citeauthoryear{{Kocevski}, {Ebeling}, {Mullis} \&
  {Tully}}{{Kocevski} et~al.}{2007}]{2007ApJ...662..224K}
{Kocevski} D.~D.,  {Ebeling} H.,  {Mullis} C.~R.,    {Tully} R.~B.,  2007,
  \apj, 662, 224

\bibitem[\protect\citeauthoryear{{Komissarov} \& {Gubanov}}{{Komissarov} \&
  {Gubanov}}{1994}]{1994A&A...285...27K}
{Komissarov} S.~S.,  {Gubanov} A.~G.,  1994, \aap, 285, 27

\bibitem[\protect\citeauthoryear{{Krymskii}}{{Krymskii}}{1977}]{1977DoSSR.234R%
1306K}
{Krymskii} G.~F.,  1977, Akademiia Nauk SSSR Doklady, 234, 1306

\bibitem[\protect\citeauthoryear{{Landau} \& {Lifshitz}}{{Landau} \&
  {Lifshitz}}{1959}]{1959flme.book.....L}
{Landau} L.~D.,  {Lifshitz} E.~M.,  1959, {Fluid mechanics}

\bibitem[\protect\citeauthoryear{{Malkov} \& {O'C Drury}}{{Malkov} \& {O'C
  Drury}}{2001}]{2001RPPh...64..429M}
{Malkov} M.~A.,  {O'C Drury} L.,  2001, Reports on Progress in Physics, 64, 429

\bibitem[\protect\citeauthoryear{{Mantz}, {Allen}, {Ebeling}, {Rapetti} \&
  {Drlica-Wagner}}{{Mantz} et~al.}{2010}]{2010MNRAS.406.1773M}
{Mantz} A.,  {Allen} S.~W.,  {Ebeling} H.,  {Rapetti} D.,    {Drlica-Wagner}
  A.,  2010, \mnras, 406, 1773

\bibitem[\protect\citeauthoryear{{Markevitch}, {Govoni}, {Brunetti} \&
  {Jerius}}{{Markevitch} et~al.}{2005}]{2005ApJ...627..733M}
{Markevitch} M.,  {Govoni} F.,  {Brunetti} G.,    {Jerius} D.,  2005, \apj,
  627, 733

\bibitem[\protect\citeauthoryear{{Mastropietro} \& {Burkert}}{{Mastropietro} \&
  {Burkert}}{2008}]{2008MNRAS.389..967M}
{Mastropietro} C.,  {Burkert} A.,  2008, \mnras, 389, 967

\bibitem[\protect\citeauthoryear{{McCarthy}, {Bower}, {Balogh}, {Voit},
  {Pearce}, {Theuns}, {Babul}, {Lacey} \& {Frenk}}{{McCarthy}
  et~al.}{2007}]{2007MNRAS.376..497M}
{McCarthy} I.~G.,  {Bower} R.~G.,  {Balogh} M.~L.,  {Voit} G.~M.,  {Pearce}
  F.~R.,  {Theuns} T.,  {Babul} A.,  {Lacey} C.~G.,    {Frenk} C.~S.,  2007,
  \mnras, 376, 497

\bibitem[\protect\citeauthoryear{{Nakazawa}, {Sarazin}, {Kawaharada},
  {Kitaguchi}, {Okuyama}, {Makishima}, {Kawano}, {Fukazawa}, {Inoue},
  {Takizawa}, {Wik}, {Finoguenov} \& {Clarke}}{{Nakazawa}
  et~al.}{2009}]{2009PASJ...61..339N}
{Nakazawa} K.,  {Sarazin} C.~L.,  {Kawaharada} M.,  {Kitaguchi} T.,  {Okuyama}
  S.,  {Makishima} K.,  {Kawano} N.,  {Fukazawa} Y.,  {Inoue} S.,  {Takizawa}
  M.,  {Wik} D.~R.,  {Finoguenov} A.,    {Clarke} T.~E.,  2009, \pasj, 61, 339

\bibitem[\protect\citeauthoryear{{Pacholczyk}}{{Pacholczyk}}{1970}]{1970ranp.b%
ook.....P}
{Pacholczyk} A.~G.,  1970, {Radio astrophysics. Nonthermal processes in
  galactic and extragalactic sources}.
Series of Books in Astronomy and Astrophysics, San Francisco: Freeman, 1970

\bibitem[\protect\citeauthoryear{{Paul}, {Iapichino}, {Miniati}, {Bagchi} \&
  {Mannheim}}{{Paul} et~al.}{2010}]{2010arXiv1001.1170P}
{Paul} S.,  {Iapichino} L.,  {Miniati} F.,  {Bagchi} J.,    {Mannheim} K.,
  2010, ArXiv e-prints

\bibitem[\protect\citeauthoryear{{Pearce}, {Thomas} \& {Couchman}}{{Pearce}
  et~al.}{1994}]{1994MNRAS.268..953P}
{Pearce} F.~R.,  {Thomas} P.~A.,    {Couchman} H.~M.~P.,  1994, \mnras, 268,
  953

\bibitem[\protect\citeauthoryear{{Peterson} \& {Fabian}}{{Peterson} \&
  {Fabian}}{2006}]{2006PhR...427....1P}
{Peterson} J.~R.,  {Fabian} A.~C.,  2006, \physrep, 427, 1

\bibitem[\protect\citeauthoryear{{Pfrommer}, {En{\ss}lin} \&
  {Springel}}{{Pfrommer} et~al.}{2008}]{2008MNRAS.385.1211P}
{Pfrommer} C.,  {En{\ss}lin} T.~A.,    {Springel} V.,  2008, \mnras, 385, 1211

\bibitem[\protect\citeauthoryear{{Poole}, {Fardal}, {Babul}, {McCarthy},
  {Quinn} \& {Wadsley}}{{Poole} et~al.}{2006}]{2006MNRAS.373..881P}
{Poole} G.~B.,  {Fardal} M.~A.,  {Babul} A.,  {McCarthy} I.~G.,  {Quinn} T.,
  {Wadsley} J.,  2006, \mnras, 373, 881

\bibitem[\protect\citeauthoryear{{Pratt} \& {Arnaud}}{{Pratt} \&
  {Arnaud}}{2002}]{2002A&A...394..375P}
{Pratt} G.~W.,  {Arnaud} M.,  2002, \aap, 394, 375

\bibitem[\protect\citeauthoryear{{Pratt}, {Croston}, {Arnaud} \&
  {B{\"o}hringer}}{{Pratt} et~al.}{2009}]{2009A&A...498..361P}
{Pratt} G.~W.,  {Croston} J.~H.,  {Arnaud} M.,    {B{\"o}hringer} H.,  2009,
  \aap, 498, 361

\bibitem[\protect\citeauthoryear{{Reiprich}, {Hudson}, {Zhang}, {Sato},
  {Ishisaki}, {Hoshino}, {Ohashi}, {Ota} \& {Fujita}}{{Reiprich}
  et~al.}{2009}]{2009A&A...501..899R}
{Reiprich} T.~H.,  {Hudson} D.~S.,  {Zhang} Y.-Y.,  {Sato} K.,  {Ishisaki} Y.,
  {Hoshino} A.,  {Ohashi} T.,  {Ota} N.,    {Fujita} Y.,  2009, \aap, 501, 899

\bibitem[\protect\citeauthoryear{{Ricker}}{{Ricker}}{1998}]{1998ApJ...496..670%
R}
{Ricker} P.~M.,  1998, \apj, 496, 670

\bibitem[\protect\citeauthoryear{{Ricker} \& {Sarazin}}{{Ricker} \&
  {Sarazin}}{2001}]{2001ApJ...561..621R}
{Ricker} P.~M.,  {Sarazin} C.~L.,  2001, \apj, 561, 621

\bibitem[\protect\citeauthoryear{{Ritchie} \& {Thomas}}{{Ritchie} \&
  {Thomas}}{2002}]{2002MNRAS.329..675R}
{Ritchie} B.~W.,  {Thomas} P.~A.,  2002, \mnras, 329, 675

\bibitem[\protect\citeauthoryear{{Roettiger}, {Burns} \& {Loken}}{{Roettiger}
  et~al.}{1993}]{1993ApJ...407L..53R}
{Roettiger} K.,  {Burns} J.,    {Loken} C.,  1993, \apjl, 407, L53

\bibitem[\protect\citeauthoryear{{Roettiger}, {Burns} \& {Stone}}{{Roettiger}
  et~al.}{1999}]{1999ApJ...518..603R}
{Roettiger} K.,  {Burns} J.~O.,    {Stone} J.~M.,  1999, \apj, 518, 603

\bibitem[\protect\citeauthoryear{{Roettiger} \& {Flores}}{{Roettiger} \&
  {Flores}}{2000}]{2000ApJ...538...92R}
{Roettiger} K.,  {Flores} R.,  2000, \apj, 538, 92

\bibitem[\protect\citeauthoryear{{Roettiger}, {Loken} \& {Burns}}{{Roettiger}
  et~al.}{1997}]{1997ApJS..109..307R}
{Roettiger} K.,  {Loken} C.,    {Burns} J.~O.,  1997, \apjs, 109, 307

\bibitem[\protect\citeauthoryear{{Roettiger}, {Stone} \& {Burns}}{{Roettiger}
  et~al.}{1999}]{1999ApJ...518..594R}
{Roettiger} K.,  {Stone} J.~M.,    {Burns} J.~O.,  1999, \apj, 518, 594

\bibitem[\protect\citeauthoryear{{Roncarelli}, {Ettori}, {Dolag}, {Moscardini},
  {Borgani} \& {Murante}}{{Roncarelli} et~al.}{2006}]{2006MNRAS.373.1339R}
{Roncarelli} M.,  {Ettori} S.,  {Dolag} K.,  {Moscardini} L.,  {Borgani} S.,
  {Murante} G.,  2006, \mnras, 373, 1339

\bibitem[\protect\citeauthoryear{{R\"ottgering}, {Wieringa}, {Hunstead} \&
  {Ekers}}{{R\"ottgering} et~al.}{1997}]{1997MNRAS.290..577R}
{R\"ottgering} H.~J.~A.,  {Wieringa} M.~H.,  {Hunstead} R.~W.,    {Ekers}
  R.~D.,  1997, \mnras, 290, 577

\bibitem[\protect\citeauthoryear{{Russell}, {Sanders}, {Fabian}, {Baum},
  {Donahue}, {Edge}, {McNamara} \& {O'Dea}}{{Russell}
  et~al.}{2010}]{2010MNRAS.406.1721R}
{Russell} H.~R.,  {Sanders} J.~S.,  {Fabian} A.~C.,  {Baum} S.~A.,  {Donahue}
  M.,  {Edge} A.~C.,  {McNamara} B.~R.,    {O'Dea} C.~P.,  2010, \mnras, 406,
  1721

\bibitem[\protect\citeauthoryear{{Russell}, {van Weeren}, {Edge}, {McNamara},
  {Sanders}, {Fabian}, {Baum}, {Canning}, {Donahue} \& {O'Dea}}{{Russell}
  et~al.}{2011}]{2011arXiv1105.0435R}
{Russell} H.~R.,  {van Weeren} R.~J.,  {Edge} A.~C.,  {McNamara} B.~R.,
  {Sanders} J.~S.,  {Fabian} A.~C.,  {Baum} S.~A.,  {Canning} R.~E.~A.,
  {Donahue} M.,    {O'Dea} C.~P.,  2011, ArXiv e-prints

\bibitem[\protect\citeauthoryear{{Sarazin}}{{Sarazin}}{2002}]{2002ASSL..272...%
.1S}
{Sarazin} C.~L.,  2002, in {L.~Feretti, I.~M.~Gioia, \& G.~Giovannini} ed.,
  Merging Processes in Galaxy Clusters Vol.~272 of Astrophysics and Space
  Science Library, {The Physics of Cluster Mergers}.
pp 1--38

\bibitem[\protect\citeauthoryear{{Schindler} \& {Mueller}}{{Schindler} \&
  {Mueller}}{1993}]{1993A&A...272..137S}
{Schindler} S.,  {Mueller} E.,  1993, \aap, 272, 137

\bibitem[\protect\citeauthoryear{{Simionescu}, {Allen}, {Mantz}, {Werner},
  {Takei}, {Morris}, {Fabian}, {Sanders}, {Nulsen}, {George} \&
  {Taylor}}{{Simionescu} et~al.}{2011}]{2011Sci...331.1576S}
{Simionescu} A.,  {Allen} S.~W.,  {Mantz} A.,  {Werner} N.,  {Takei} Y.,
  {Morris} R.~G.,  {Fabian} A.~C.,  {Sanders} J.~S.,  {Nulsen} P.~E.~J.,
  {George} M.~R.,    {Taylor} G.~B.,  2011, Science, 331, 1576

\bibitem[\protect\citeauthoryear{{Skillman}, {Hallman}, {O'Shea}, {Burns},
  {Smith} \& {Turk}}{{Skillman} et~al.}{2011}]{2011ApJ...735...96S}
{Skillman} S.~W.,  {Hallman} E.~J.,  {O'Shea} B.~W.,  {Burns} J.~O.,  {Smith}
  B.~D.,    {Turk} M.~J.,  2011, \apj, 735, 96

\bibitem[\protect\citeauthoryear{{Springel} \& {Farrar}}{{Springel} \&
  {Farrar}}{2007}]{2007MNRAS.380..911S}
{Springel} V.,  {Farrar} G.~R.,  2007, \mnras, 380, 911

\bibitem[\protect\citeauthoryear{{Takizawa}}{{Takizawa}}{1999}]{1999ApJ...520.%
.514T}
{Takizawa} M.,  1999, \apj, 520, 514

\bibitem[\protect\citeauthoryear{{Takizawa}}{{Takizawa}}{2000}]{2000ApJ...532.%
.183T}
{Takizawa} M.,  2000, \apj, 532, 183

\bibitem[\protect\citeauthoryear{{Takizawa}}{{Takizawa}}{2006}]{2006PASJ...58.%
.925T}
{Takizawa} M.,  2006, \pasj, 58, 925

\bibitem[\protect\citeauthoryear{{Takizawa}}{{Takizawa}}{2008}]{2008ApJ...687.%
.951T}
{Takizawa} M.,  2008, \apj, 687, 951

\bibitem[\protect\citeauthoryear{{Takizawa} \& {Naito}}{{Takizawa} \&
  {Naito}}{2000}]{2000ApJ...535..586T}
{Takizawa} M.,  {Naito} T.,  2000, \apj, 535, 586

\bibitem[\protect\citeauthoryear{{van Weeren}, {Hoeft}, {R{\"o}ttgering},
  {Br{\"u}ggen}, {Intema} \& {van Velzen}}{{van Weeren}
  et~al.}{2011}]{2011A&A...528A..38V}
{van Weeren} R.~J.,  {Hoeft} M.,  {R{\"o}ttgering} H.~J.~A.,  {Br{\"u}ggen} M.,
   {Intema} H.~T.,    {van Velzen} S.,  2011, \aap, 528, A38+

\bibitem[\protect\citeauthoryear{{van Weeren}, {Intema}, {Rottgering},
  {Bruggen} \& {Hoeft}}{{van Weeren} et~al.}{2011}]{2011arXiv1101.5161V}
{van Weeren} R.~J.,  {Intema} H.~T.,  {Rottgering} H.~J.~A.,  {Bruggen} M.,
  {Hoeft} M.,  2011, ArXiv e-prints

\bibitem[\protect\citeauthoryear{{van Weeren}, {R{\"o}ttgering}, {Bagchi},
  {Raychaudhury}, {Intema}, {Miniati}, {En{\ss}lin}, {Markevitch} \&
  {Erben}}{{van Weeren} et~al.}{2009}]{2009A&A...506.1083V}
{van Weeren} R.~J.,  {R{\"o}ttgering} H.~J.~A.,  {Bagchi} J.,  {Raychaudhury}
  S.,  {Intema} H.~T.,  {Miniati} F.,  {En{\ss}lin} T.~A.,  {Markevitch} M.,
  {Erben} T.,  2009, \aap, 506, 1083

\bibitem[\protect\citeauthoryear{{van Weeren}, {R{\"o}ttgering} \&
  {Br{\"u}ggen}}{{van Weeren} et~al.}{2011}]{2011A&A...527A.114V}
{van Weeren} R.~J.,  {R{\"o}ttgering} H.~J.~A.,    {Br{\"u}ggen} M.,  2011,
  \aap, 527, A114+

\bibitem[\protect\citeauthoryear{{van Weeren}, {R{\"o}ttgering}, {Br{\"u}ggen}
  \& {Cohen}}{{van Weeren} et~al.}{2009}]{2009A&A...508...75V}
{van Weeren} R.~J.,  {R{\"o}ttgering} H.~J.~A.,  {Br{\"u}ggen} M.,    {Cohen}
  A.,  2009, \aap, 508, 75

\bibitem[\protect\citeauthoryear{{van Weeren}, {R{\"o}ttgering}, {Br{\"u}ggen}
  \& {Hoeft}}{{van Weeren} et~al.}{2010}]{2010Sci...330..347V}
{van Weeren} R.~J.,  {R{\"o}ttgering} H.~J.~A.,  {Br{\"u}ggen} M.,    {Hoeft}
  M.,  2010, Science, 330, 347

\bibitem[\protect\citeauthoryear{{Venturi}, {Giacintucci}, {Brunetti},
  {Cassano}, {Bardelli}, {Dallacasa} \& {Setti}}{{Venturi}
  et~al.}{2007}]{2007A&A...463..937V}
{Venturi} T.,  {Giacintucci} S.,  {Brunetti} G.,  {Cassano} R.,  {Bardelli} S.,
   {Dallacasa} D.,    {Setti} G.,  2007, \aap, 463, 937

\bibitem[\protect\citeauthoryear{{Vikhlinin}, {Kravtsov}, {Forman}, {Jones},
  {Markevitch}, {Murray} \& {Van Speybroeck}}{{Vikhlinin}
  et~al.}{2006}]{2006ApJ...640..691V}
{Vikhlinin} A.,  {Kravtsov} A.,  {Forman} W.,  {Jones} C.,  {Markevitch} M.,
  {Murray} S.~S.,    {Van Speybroeck} L.,  2006, \apj, 640, 691

\bibitem[\protect\citeauthoryear{{Voges}, {Aschenbach}, {Boller},
  {Br{\"a}uninger}, {Briel}, {Burkert}, {Dennerl} \& {et al.}}{{Voges}
  et~al.}{1999}]{1999A&A...349..389V}
{Voges} W.,  {Aschenbach} B.,  {Boller} T.,  {Br{\"a}uninger} H.,  {Briel} U.,
  {Burkert} W.,  {Dennerl} K.,    {et al.} 1999, \aap, 349, 389

\bibitem[\protect\citeauthoryear{{Zu Hone}, {Ricker}, {Lamb} \& {Karen
  Yang}}{{Zu Hone} et~al.}{2009}]{2009ApJ...699.1004Z}
{Zu Hone} J.~A.,  {Ricker} P.~M.,  {Lamb} D.~Q.,    {Karen Yang} H.-Y.,  2009,
  \apj, 699, 1004

\bibitem[\protect\citeauthoryear{{ZuHone}}{{ZuHone}}{2011}]{2011ApJ...728...54%
Z}
{ZuHone} J.~A.,  2011, \apj, 728, 54

\bibitem[\protect\citeauthoryear{{ZuHone}, {Markevitch} \& {Johnson}}{{ZuHone}
  et~al.}{2010}]{2010ApJ...717..908Z}
{ZuHone} J.~A.,  {Markevitch} M.,    {Johnson} R.~E.,  2010, \apj, 717, 908

\end{thebibliography}

\label{lastpage}

\end{document}